\documentclass[12pt]{article}
\usepackage{amsmath,mathdots}
\usepackage{amsfonts}
\usepackage{tikz}
\usepackage{latexsym}
\usepackage{graphicx}
\usepackage{color}
\usepackage{xcolor}
\usepackage{amsmath, amssymb,latexsym}
\usepackage{color}
\usepackage{fullpage}
\usepackage[figurewithin=none]{caption}

\usepackage{fullpage,graphicx}
\usepackage{rotating}

\usetikzlibrary{positioning}
\usetikzlibrary{shapes,arrows}

\DeclareMathOperator{\E}{\mathbb{E}}

\newtheorem{theorem}{Theorem}[section]

\title{Portfolio Management Using Graph Centralities:\\ 
Review and Comparison}
\author{Bahar Arslan \thanks{Bursa Technical University,
        Mathematics Department, Bursa, Turkey}
\and
 Vanni Noferini \thanks{Aalto University, Department of Mathematics and Systems Analysis, P.O. Box 11100, FI-00076, Aalto, Finland. Supported by an Academy of Finland grant (Suomen Akatemian p\"{a}\"{a}t\"{o}s 331240).}
 \and  
 Spyridon Vrontos \thanks{School of Mathematics, Statistics and Actuarial Science, University of Essex, United Kingdom} }
\begin{document}
\maketitle

\begin{abstract}
We investigate an application of network centrality measures to portfolio optimization, by generalizing the method in [Pozzi, Di Matteo and Aste, \emph{Spread of risks across financial markets: better to invest in the peripheries}, Scientific Reports 3:1665, 2013], that however had significant limitations with respect to the state of the art in network theory. In this paper, we systematically compare many possible variants of the originally proposed method on S\&P 500 stocks. We use daily data from twenty-seven years as training set and their following year as test set. We thus select the best network-based methods according to different viewpoints including for instance the highest Sharpe Ratio and the highest expected return. We give emphasis in  new centrality measures and we also conduct a thorough analysis, which reveals significantly stronger results compared to those with more traditional methods.
According to our analysis, this graph-theoretical approach to investment can be used successfully by investors with different investment profiles leading to high risk-adjusted returns.
\end{abstract}

\section{Introduction}
In graph theory and network theory, a centrality measure is a nonnegative valued function defined on each node of a graph \cite{estrada,EH,newman}. The goal of such measures is to characterize what are, in some sense, the ``most important'' vertices in a network; this is a question of interest across various field: to name but a few, biology, medicine, physics, and social sciences. Examples of applications include identifying relevant web pages, influential users in a social network, or superspreaders of diseases.

Recently, the application of these ideas to finance was proposed in \cite{PMA13}, in the context of portfolio optimization. There, a number of centrality measures are considered, but with some significant limitation with respect to the possibilities available in the literature. For example, Katz centrality which is arguably the most popular among walk-based centrality measures, depends on a parameter $\alpha$: \cite{PMA13} only investigates the two extreme cases of degree centrality (which is Katz centrality in the limit of small $\alpha$) and eigenvector centrality (which is Katz centrality in the limit of large $\alpha$), neglecting the possibilities in between. Another possible limitation of the study in \cite{PMA13} is that the authors are only retaining, in every market day, the top 300 best performing stocks out of the 600 with highest capitalization. With this method, the actual predicting power of centrality measures might be hidden by its enhancement through such a momentum stocks selection procedure.

In this paper, we aim to fill these gaps by proposing a more systematic comparison of portfolio selection strategies purely based on graph centrality, including a study of potentially different behaviours when a parameter, such as Katz's $\alpha$, varies within its range of possible values. Moreover, we include in our analysis more recent and mathematically sophisticated measures that have been shown to be qualitatively different than its predecessors, such as NBTW centrality \cite{AGHN17, AGHN18,GHN18}. We have compared, and tested on historical data from the stocks included in the S\&P 500 index, various centrality measures (possibly depending on a parameter $\alpha$) as well as possible associated tweaks such as how to exactly construct the underlying adjacency matrix, whether to look at central or peripheral nodes, and how to precisely use the centrality measure in building the portfolio. We note that, although perhaps not as thoroughly, the latter two aspects are studied already in \cite{PMA13}, whereas the former seems to be investigated here for the first time.

We propose the use of a much wider range of centrality measures, including the fairly new NBTW and the use of subgraph centralities (Exponential, Katz and NBTW) which perform excellently on our tests. In addition, we propose the use of a plethora of transformations of correlations matrices and alternative formulations for the adjacency matrices. We also utilize a simple threshold instead of other more sophisticated filtering techniques and we depict the most frequently used threshold values. 

\section{Preliminaries}
In this section we discuss the necessary background in network theory and in financial mathematics.

\subsection{Centrality measures}

Recall that an \emph{undirected graph} is a set of nodes together with a set of (unordered) pairs of nodes, called \emph{edges}. Pictorially, edges pairwise connect nodes. The edges may or may not be \emph{weighted} by positive real numbers; in an unweighted graph, the weights are all equal to $1$. Given a labelling of the nodes in a graph from $1$ to $n$, the \emph{adjacency matrix} of the graph is the $n \times n$ matrix whose $(i,j)$ is equal to the weight of the edge between node $i$ and node $j$, or equal to $0$ if no edge is present. Adjacency matrices of unweighted graphs have entries lying in $\{0,1\}$, while adjacency matrices of weighted graph can have more general nonnegative real numbers as entries; relabelling the nodes corresponds to performing a permutation similarity on the adjacency matrix, and in particular, it leaves all the eigenvalues unchanged. A graph is called simple if no loops (edges from a node to itself) are allowed. The diagonal elements of the adjacency matrix of a simple graph are all zero, and conversely, if at least one diagonal element of the adjacency matrix of a graph is nonzero then the graph is not simple.

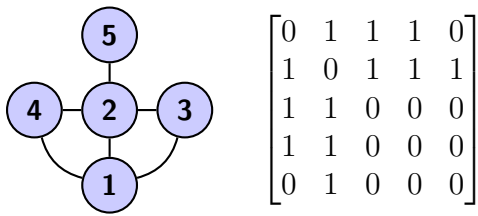
\begin{figure}[h!]
\begin{center}
\begin{tikzpicture}[ >=stealth',shorten >=0pt,auto,node distance=1.0cm,
  thick,main node/.style={circle,fill=blue!20,draw,font=\sffamily\small\bfseries}]

  \node[main node] (5){5};
  \node[main node](2) [below of=5]{2};
  \node[main node] (3) [right of=2] {3};
  \node[main node](4) [left of=2]{4};
  \node[main node](1) [below of=2]{1};
 \node[](6) at(3.5,-1) {$ \begin{bmatrix}
0 & 1 & 1 & 1 & 0\\
1 & 0 & 1 & 1 & 1\\
1 & 1 & 0 & 0 & 0\\
1 & 1 & 0 & 0 & 0\\
0 & 1 & 0 & 0 & 0
\end{bmatrix}$};

  \path[every node/.style={font=\sffamily\small}]
            (3) edge node [right]{} (2)
    (4) edge node [right] {} (2)
    (5) edge node [right] {} (2)
       (2) edge [right] node[left] {} (1)
       (3) edge [bend left] node[left] {} (1)
        (4) edge [bend right] node[left] {} (1);
\end{tikzpicture}

\end{center} 
\caption{Left: A simple, unweighted, undirected graph with 5 nodes (left) whose edges are, following the same labelling as in the picture, $\{1,2\}, \{1,3\}, \{1,4\}, \{2,3\}, \{2,4\},$ and $\{2,5\}$. Right: The adjacency matrix of the graph on the left, again following the same labelling of the nodes.}
\end{figure}

Mathematically, a \emph{centrality measure} is a nonnegative function defined on each node. The significance of centrality measures in network science is that they are defined in such a way that they can somehow capture the relative importance, or centrality, of each node; another feature taken into account in the definition of centrality measure is the ease of their numerical computation. Many centrality measures are based on the combinatorics of \emph{walks}: it is therefore useful to recall that a walk is a finite sequence of nodes such that any two consecutive nodes in a walk are connected by an edge.

In this section we briefly review the centrality measures that we use, describing their features and touching on their mathematical justification. We use the notation $A$ to denote the adjacency matrix of the graph, $I=A^0$ is the identity matrix of the same size, and $\mathbf{1}$ to denote the vector of all ones.

In the network theory literature, there are hundreds of centralities that have been proposed for applications. We wanted to test a wide enough range of methods, while at the same time selecting the most significant ones to avoid overfitting. The criteria that we used the select the most relevant centralities are:
\begin{enumerate}
\item \emph{Relevance.} Frequency of use in the specialist literature, and impact in this field.
\item \emph{Computational efficiency.} Although the typical size of the graphs encountered in practice in portfolio selection is very small (in the order of hundreds or at most thousands of nodes), and thus at first sight one may think that efficiency of computations is irrelevant, as we will see below the computation of a centrality measure must be repeted many times. Moreover, it is conceivable that practitioners in finance may wish to use some of these methods in a high frequency trading context, and hence, they may also value the ability of not wasting even the tiniest fraction of a second. For this reason, we favoured methods that are efficient both in terms of theoretical complexity and in terms of efficient practical implementation (in particular, methods based on numerical linear algebra are typically extremely fast in practice, being able to exploit the very well maintained existing software libraries).
\item \emph{Success.} Amount of evidence we could find in the literature that supported the ability of a centrality to provide insight on the underlying phenomena modelled by a graph, both within finance and in other fields.
\item \emph{Potential and diversity.} Not to exclude more recently proposed centralities, we took into account the potential predicted power of newer ideas, again based on the evidence available in the literature. In particular, we included ideas that are significantly different than predecessors, and we neglected ideas that are too similar to other already included methods.
\end{enumerate}

\subsubsection{Degree centrality}

One of the simplest centrality measures, the degree centrailty of each node is defined to be the node's \emph{degree}, i.e., the number of edges connected to it. For an unweighted graph, the vector of degree centralities can be computed  as the diagonal elements of $D$ where $D$ is the diagonal matrix of degrees.

\subsubsection{Katz centrality}

First proposed by the social scientist L. Katz, Katz centrality is based on the combinatorics of walks. Its value on node $i$ is the weighted sum of all the walks starting from node $i$, where walks of length $k$ are weighted $t^k$ where $\alpha$ is a real parameter in $(0,\rho(A)^{-1})$ and $\rho(A) \geq 1$ is the spectral radius of $A$, or equivalently since $A$ is symmetric and nonnegative, its largest eigenvalue. Katz centrality depends on the choice of the parameter $\alpha$ and can be computed solving a linear system:
$$ \mathbf{v} = (I - \alpha A)^{-1} \mathbf{1}.$$
When $t \rightarrow 0$, Katz centrality tends to degree centrality, while the limit $t \rightarrow \rho(A)^{-1}$ is known as \emph{eigenvector centrality}.

\subsubsection{Subgraph centrality}

A variant of Katz centrality, in which rather than counting all the walks starting from node $i$ one counts only closed walks (i.e. walks starting from and ending on node $i$). It still depends on the same parameter $\alpha$ as Katz centrality, and it can be computed as the diagonal of the matrix $(I-tA)^{-1}$.

\subsubsection{Exponential centrality}

It is defined similarly to Katz centrality, but the weight of walks of lenght $k$ is smaller and equal to $\alpha^k/k!$. The range of possible values of the parameter $\alpha$ is also different and equal to $(0,\infty)$. Exponential centrality can be computed as
$$\mathbf{v} = e^{\alpha A} \mathbf{1}.$$

\subsubsection{Exponential subgraph centrality}

The exponential variant of subgraph centrality can be computed as the diagonal of the matrix $e^{\alpha A}$.

\subsubsection{NBTW centrality}

The recently introduced NBTW centrality \cite{AGHN18,GHN18} is constructed in the same way as Katz, but only counting \emph{nonbacktracking walks} (NBTWs), i.e., walks that do not contain any subsequence of nodes of the form $iji$. It depends on a parameter $\alpha$ whose range of values is $(0,\mu)$ and $\mu > \rho(A)^{-1}$, i.e., the range is always larger than the one for Katz. For unweighted undirected graphs without loops, it can be computed solving a linear system \cite{GHN18}:
$$\mathbf{v} = (1-\alpha^2) (I - \alpha A + \alpha^2(D-I) )^{-1} \mathbf{1}.$$
The limit of NBTW centrality when $\alpha \rightarrow 0$ is again degree centrality, while the limit when $\alpha \rightarrow \mu$ is known as \emph{nonbacktracking eigenvector centrality}, and it is generally not the same as (classical) eigenvector centrality: see \cite{GHN18}. For weighted graphs, the computation of NBTW centrality still comes by solving a linear system, but the construction of the latter is more complicated \cite{Ryan}. In \cite{Ryan}, an assumption was made that the graph had no loops. One can also consider NBTWs with loops: since a loop is a subsequence of the form $ii$, this can trigger backtracking in a walk if the same loop is traversed twice in a row. It turns out that the formulae obtained in \cite{Ryan}, that also cover those in \cite{AGHN18,GHN18} as special cases, are still valid even if loops are present. For completeness, we make a formal statement below; the proof is essentially the same as in \cite{Ryan}, but we sketch it for completeness.

\begin{theorem}\label{thm:withloops}
    Let $G=(V,E)$ be a finite graph with $n$ nodes, possibly directed, possibly weighted and possibly with loops; let $A \in \mathbb{R}^{n \times n}$ be the adjacency matrix of $G$, i.e., $A_{ij}>0$ is equal to the weight of the edge $(i,j)$ if $(i,j) \in E$ is an edge of $G$, and $A_{ij}=0$ if $(i,j) \not \in E$. Define the rational matrix $\Psi(\alpha)=\Psi_e(\alpha)-\Psi_o(\alpha)$, $\Psi_e(\alpha)$ is diagonal, and
\[  (\Psi_e(\alpha))_{ii}=  1 + \sum_{j=1}^n \frac{\alpha^2 A_{ij} A_{ji}}{1-\alpha^2 A_{ij} A_{ji}}, \qquad (\Psi_o(\alpha))_{ij}=\frac{\alpha A_{ij}}{1-\alpha^2 A_{ij}A_{ji}}    .\]
Then, $\Psi(\alpha)^{-1}$ generates the weighted enumeration of NBTWs on $G$, that is, it has the Taylor expansion $\Psi(\alpha)^{-1} = \sum_{k=0}^\infty \alpha^k P_k$, and $(P_k)_{ij} \geq 0$ is the weighted sum of all NBTWs of length $k$ from $i$ to $j$ on $G$, with weight given by the product of the weights of all the $k$ edges traversed in the walk.
\end{theorem}

Before proving Theorem \ref{thm:withloops}, we remark that with respect to the case without loops \cite{Ryan}, $\Psi_e(\alpha)$ is still a diagonal matrix but $\Psi_o(\alpha)$ is no longer the off-diagonal part of $\Psi(\alpha)$, because not necessarily $A_{ii}=0$ for all $i$.

\begin{proof}[Proof of Theorem \ref{thm:withloops}]
   Following \cite[Theorem 3.1]{Ryan}, the first step in the proof is deriving a recurrence for the matrix coefficients $P_k$ in $\Psi(\alpha)=\sum_{k=0} P_k \alpha^k$. This is done by noting that $(A P_{k-1})_{ij}$ is equal to the weighted enumeration of all walks of length $k$ from $i$ to $j$ that become nonbacktracking if the initial step is removed. In turn this is equal to $(P_k)_{ij}$ plus the weighted enumeration of all \emph{backtracking} walks of length $k$ from $i$ to $j$ that become nonbacktracking if the initial step is removed. The walks that contribute to such discrepancy must have the form $iai \dots j$; the difference with respect to the case without loops in \cite{Ryan} is simply that we now possibly allow $a=i$, but the formula for their summation over $a$ is formally the same, namely, $\mathrm{dd}(A^2)P_{k-2}$ where $\mathrm{dd}(X)$ denote a diagonal matrix whose diagonal elements coincide with those of the matrix $X$. However, we have now subtracted too much, because walks of the form $iaia \dots j$ were not counted in $(A P_{k-1})_{ij}$, as they were already backtracking even after removing the first step. One thus needs to recursively correct the combinatorial enumeration until the length of the walk is exhausted. This eventually leads to the same recursion as in \cite{Ryan}, with the only difference that the diagonal elements of $A$ may be nonzero and thus contribute to the count, but without formally changing the formulae.

We thus obtain a recurrence of the form $\sum_{\ell=0}^k C_\ell P_{k - \ell}=0$ for all $k \geq 1$, where $C_k$ are also matrices that can be obtained explicitly and are as in \cite[Theorem 3.1]{Ryan}; we omit their precise explicit expression as it is not crucial to describe the argument, as long as it is explicitly known. Moreover, by convention $C_0=P_0=I$. Thanks to \cite[Proposition 3.2]{Ryan}, this implies that the generating function are inverses of each other, i.e, $\sum_{k=0}^\infty P_k \alpha^k = \left(\sum_{k=0}^\infty C_k \alpha^k\right)^{-1}.$  At this point, obtaining the expression of $\Psi(\alpha)$ from that of the $C_k$ is elementary and can be done following verbatim the arguments leading to \cite[Theorem 3.3]{Ryan}.
\end{proof}

Once $\Psi(\alpha)$ is known, a formula for NBTW centrality measures follows as a corollary because it is defined to be $\mathbf{v}=\Psi(\alpha)^{-1} \mathbf{1}$; this is again tantamount to solving a linear system with coefficient $\Psi(\alpha)$ and right hand side $\mathbf{1}$. Note, however, that Theorem \ref{thm:withloops} considers possibly directed graphs, but in this paper we are only interested in undirected ones. This simplies somewhat the formula, because for undirect graphs $A_{ij}=A_{ji}$. More explicitly, in the case of our interest the matrix $\Psi(\alpha)$ is as follows:
\begin{equation*}\label{eq:ryan}
\Psi(\alpha)_{ij} = \begin{cases} 1-\frac{\alpha A_{ii}}{1-\alpha^2 A_{ii}}+\sum_{j=1}^n \frac{\alpha^2 A_{ii}^2}{1-\alpha^2 A_{ij}^2} \ &\mathrm{if} \ i=j;\\
-\frac{\alpha A_{ij}}{1-\alpha^2 A_{ij}} \ &\mathrm{if} \ i \neq j.
\end{cases}
\end{equation*}

\subsubsection{NBTW subgraph centrality}

The nonbacktracking analogue of subgraph centrality can be computed, in the unweighted case \cite{GHN18}, as the diagonal elements of the matrix $(1-\alpha^2) (I - \alpha A + \alpha^2(D-I) )^{-1}$ or, in the weighted case \cite{Ryan}, as the diagonal elements of the matrix $\Psi(\alpha)^{-1}$.

\subsubsection{NBTW exponential centrality}
The exponential variants of NBTW centrality and NBTW subgraph centrality can be computed in a mathematically more complicated, but still computationally efficient, way exploting certain matrix functions. Interested readers can refer to \cite{AGHN18,Ryan} for more details on this method. Note that, while \cite[Theorem 4.6]{Ryan} assumes that the graphs are loopless, once again it is not difficult to extend the results therein by showing that they also hold in the case with loops (with the same formal expressions).

\subsubsection{NBTW exponential subgraph centrality}

We again refer to \cite{AGHN18,Ryan}, and in particular to \cite[Theorem 4.6]{Ryan} for the details of the method we used to compute this centrality.

\subsubsection{Betweenness centrality}

Betweenness centrality is based on \emph{paths}: walks whose nodes are all distinct except possibly the first and the last ones. In particular, its value on node $i$ is computed as the sum, over all pairs of nodes $(j,k)$ with $j \neq i \neq k$, of the fraction of shortest paths between node $j$ and node $k$ that pass through node $i$.

\subsection{Portfolio optimization}
We consider an investor who allocates their wealth among the $m$ (most central or most peripheral) stocks 
with portfolio weight vector ${\bf x}= \begin{bmatrix} x_{1} & x_{2} & \dots & x_{m}\end{bmatrix}^{T}$ and 
proceeds in re-balancing their portfolio every $\Delta t$ trading days.

We employ three different approaches to construct the optimal portfolios:
\begin{enumerate}
	\item Equal Weights: 
	 \[x_{1}=x_{2}=\cdots=x_{m}=\frac{1}{m}.\]

	\item Minimum Variance:
	\begin{align*}
		\min_{\bf x} \quad &  \dfrac{1}{2}{\bf x}^T\Sigma {\bf x}; \\
		s.t. \quad & {\bf x}^T \mathbf{1} = 1,
\end{align*}
where $\Sigma$ is the covariance matrix. 
We consider both options regarding short selling, having $0 \leq x_i \leq 1$ $\forall$ $i$ in case that short selling is not allowed and 
$-1 \leq x_i \leq 1$ $\forall$ $i$ in case that short selling is allowed. Note that ${\bf x}^T {\bf 1} = \| {\bf x} \|_1$ in the former case, but not in the latter.

\item Mean-Variance:
\begin{align*}
\min_{\bf x} \quad & \dfrac{1}{2} {\bf x}^T\Sigma {\bf x} \\
s.t. \quad & {\bf x}^T \mathbf{1}= 1 \\
&{\bf x}^T {\bf r}= r_{target}
\end{align*}
where ${\bf r}$ is the vector of returns and $r_{target}$ is the target mean return. 
\end{enumerate}
 We set an annual target return, $r_{target}$, in the optimization schemes used.
The target return is not always achieved by the optimization procedure; in these cases, the target is lowered to the highest
possible positive value. 
For more about portfolio optimization see, e.g., \cite{EG,Markowitz,Meucci}. For more on the comparison of portfolios of equal weights with other more advanced portfolios we refer to \cite{DeMiguelEtAl}.

\subsection{Performance evaluation criteria}
The performance of the constructed portfolios is evaluated over a long out-of-sample period of twenty-seven years using a plethora of performance measures. 
First, we consider the realized returns of the constructed portfolios. Given
the portfolio weights at time $t$, denoted by ${\bf x}_{t}$, and the realized
returns of the $m$ assets in our sample at time $t+1$, denoted by ${\bf r}_{t+1}= \begin{bmatrix} r_{1} & r_{2} & \dots & r_{m}\end{bmatrix} _{t+1}^{T }$, the realized
return $r_{p}$ of the portfolio at time $t+1$ is computed as%
\begin{equation*}
r_{p,t+1}={\bf x}_{t}^{T} {\bf r}_{t+1}.
\end{equation*}%
We calculate the Expected Return (ER) within the out-of-sample period, 
the cumulative return at the end of the period, and the volatility (SD) of the portfolio.  Due to the fact that portfolio optimization
schemes generally arrive at a different minimum variance for each network/correlation/centrality structure, the realized return is not comparable across models since it represents portfolios bearing different risks. Therefore, a more appropriate/realistic approach is to compare the return per unit of risk. In this sense, we use the Sharpe Ratio (SR) which standardizes the realized
returns with the risk of the portfolio and is calculated through%
\begin{equation*}
SR_{p}=\frac{\E[r_{p}]-\E[r_{f}]}{\sqrt{Var(r_{p})}},
\end{equation*}%
where $r_{p}$ is the average realized return of the portfolio over the
out-of-sample period, and $Var(r_{p})$ is the variance of the portfolio over
the out-of-sample period$.$

A portfolio measure associated with the sustainability of the portfolio
losses is the maximum drawdown (MDD) which broadly reflects the maximum
cumulative loss from a peak to a following bottom. MDD is defined as the
maximum sustained percentage decline (peak to trough), which has occurred in
the portfolio within the period studied and is calculated from the following
formula

\begin{equation*}
MDD_{p}=\max_{T_{0}\leq t\leq T-1}[\max_{T_{0}\leq j\leq
T-1}(PV_{j})-PV_{t}],
\end{equation*}%
where $PV$ denotes the portfolio value and $T_{0},T$ denote the beginning
and end of the evaluation period, respectively.

The next three measures that we calculate, namely the Omega (OMG), Sortino (SOR)
and Upside Potential (UP) ratios, treat portfolio losses and gains
separately. In order to define these measures, we first introduce the positive part function of a real variable, as
$$x_+ = \frac{x + |x|}{2}.$$ Next, we define the n-th
lower partial moment ($LPM_{n})$ of the portfolio return as follows (for more details, see \cite{Bacon}):
\begin{equation*}
LPM_{n}(r_{b})=\E[((r_{b}-r_{p})_{+})^{n}]
\end{equation*}%
where $r_{b}$ is the benchmark return and the Kappa function ($K_{n}(r_{b}))$
is defined as follows:

\begin{equation*}
K_{n}(r_{b})=\frac{\E(r_{p})-r_{b}}{\sqrt[n]{LPM_{n}(r_{b})}}.
\end{equation*}%
Then the respective measures are calculated as follows: 
\begin{equation*}
OMG(r_{b})=K_{1}(r_{b})+1, \qquad
SOR(r_{b})=K_{2}(r_{b}), \qquad
UP(r_{b})=\frac{%
\E[(r_{p}-r_{b})_{+}]}{\sqrt{LMPM_{2}(r_{b})}}.
\end{equation*}

Finally, we investigate the tail-risk of the different models proposed.
A CVaR of $\lambda $\% at the 100(1-$\alpha $)\%
confidence level means that the average portfolio loss measured over 100$%
\alpha $\% of worst cases is equal to $\lambda $\% of the wealth managed by
the investor. To compute CVaR, we use the empirical distribution of
the portfolio realized returns. CVaR is calculated at the 95\% confidence level.

We employ the US 3-month interest rate for the risk free rate and for the
benchmark rate of return $(r_{b})$ necessary for the calculation of SR, OMG, MDD
and SOR. 

\section{Methodology}

In this section we describe in detail the exact implementation of the methods we have employed.

\subsection{Constructing the correlation matrix from the data}

The first step in all of our methods is to build the raw correlation matrix $C$. This will be then used in various different way to build an adjacency matrix $A$ (or, equivalently, a graph).

Our method for this preliminary step follows \cite{PMA13}.  We start from the historical returns of the stocks in the S\&P 500 index, recorded daily over the years 1990-2017. For each year, we use a moving window of $\tau = 125$ days; we calculate Pearson's correlation coefficients over such a window, ending $t$ days before the end of the year for $t=1,\dots,\tau.$ These correlations are then summed, with weights according to the formula
$$w(t) = w_0 \exp \left( \frac{t-\tau}{\tau} \right),$$
where the normalization $w_0$ satisfies $\sum_{t=1}^\tau w(t)=1.$

A possible next preprocessing step, that we take as optional in order to analyse its effect, is what is called \emph{shrinkage} \cite{LW03} of the correlation matrix $C$. This consists in constructing the covariance matrix from the correlation matrix; then, a linear combination of the covariance matrix and a matrix coming from a single-index model is computed; finally, one divides by the standard deviations again. For more details, see \cite{LW03}. In the following, we refer to the correlation matrix constructed as described above (with or without shrinkage) as $C$.

\subsection{Constructing the graph from the correlation matrix}

The next stage is to construct a (weighted or unweighted) graph, or equivalently its adjacency matrix $A$, given the correlation matrix $C$. We kept various options open on how to do it, which we overview below.

To illustrate each method, we will follow how the toy example of the correlation matrix (to four decimal points)

\begin{equation*}\label{eq:toyexample} C = \begin{bmatrix}
 1  & -0.1378 &   0.2025 &   0.4683  & -0.2583\\
   -0.1378 &   1 &   0.4373  &  0.1050 &   -0.1738\\
   0.2025  &  0.4373 &   1 &   0.4245 &   0.4108\\
    0.4683  &  0.1050  &  0.4245   & 1  & -0.0465\\
   -0.2583 &  -0.1738  &  0.4108  & -0.0465  &  1
\end{bmatrix} \end{equation*}
will be processed to generate an adjacency matrix $A$.

\subsubsection{Correlation matrix with loops}

The simplest option is to take $A=|C|$ (where the absolute value function is applied element-wise). This way, the underlying graph has weights given by the absolute value of the correlation between any pair of stocks. The rationale is that we are interested in investigating how stocks influence each other's behaviour regardless of whether the correlation is positive or negative, but only looking at its strength.

 Usually, a (weighted) complete graph is generated with this method. We also note that such a graph has loops, i.e., edges going from one node to itself, with weight $1$ since $C_{ii}=1$.
 
 For \eqref{eq:toyexample} this method generates
 $$ A = \begin{bmatrix}
 1  & 0.1378 &   0.2025 &   0.4683  & 0.2583\\
   0.1378 &   1 &   0.4373  &  0.1050 &   0.1738\\
   0.2025  &  0.4373 &   1 &   0.4245 &   0.4108\\
    0.4683  &  0.1050  &  0.4245   & 1  & 0.0465\\
   0.2583 &  0.1738  &  0.4108  & 0.0465  &  1
\end{bmatrix}$$

 \subsubsection{Correlation matrix without loops}
 
 If one wants a simple graph, i.e., loops are to be avoided, the previous method can be simply modified as $A=|C|-I.$
 
 For \eqref{eq:toyexample} one gets
  $$ A = \begin{bmatrix}
 0  & 0.1378 &   0.2025 &   0.4683  & 0.2583\\
   0.1378 &   0 &   0.4373  &  0.1050 &   0.1738\\
   0.2025  &  0.4373 &   0 &   0.4245 &   0.4108\\
    0.4683  &  0.1050  &  0.4245   & 0  & 0.0465\\
   0.2583 &  0.1738  &  0.4108  & 0.0465  &  0
\end{bmatrix}$$
 \subsubsection{Unweighted thresholded correlation matrix}
 
 It is possible to construct a graph other than the complete one by thresholding the correlation matrix. For example, one may want to only keep positive correlations ignoring the negative ones. Or one could still adopt the approach of only looking at the absolute value of the correlations, still mapping low absolute values to zero. Again, there is also an option of keeping the loops or not. 
 
 Mathematically, this is realized in the following way: we define the logical function $x>y$ to be equal to $1$ if $x$ is effectively greater than $y$ and $0$ otherwise. We then choose a threshold $\theta \in (0,1)$. Finally, we apply the scalar function above elementwise to matrices to obtain the formulae
  \begin{eqnarray}
 A &=& C > \theta, \label{loopsnotransformunweighted}\\
 A &=& |C| > \theta, \label{loopsabsunweighted}\\
 A &=& (C - I) > \theta, \label{noloopsnotransformunweighted}\\ 
 A &=& (|C|-I) > \theta. \label{noloopsabsunweighted}
 \end{eqnarray}
This approach always builds an unweighted graph. Indeed, $A$ is a matrix whose entires belong to $\{0,1\}$. The graph is simple if and only if we subtract the identiy matrix $I$ from $C$ before thresholding.

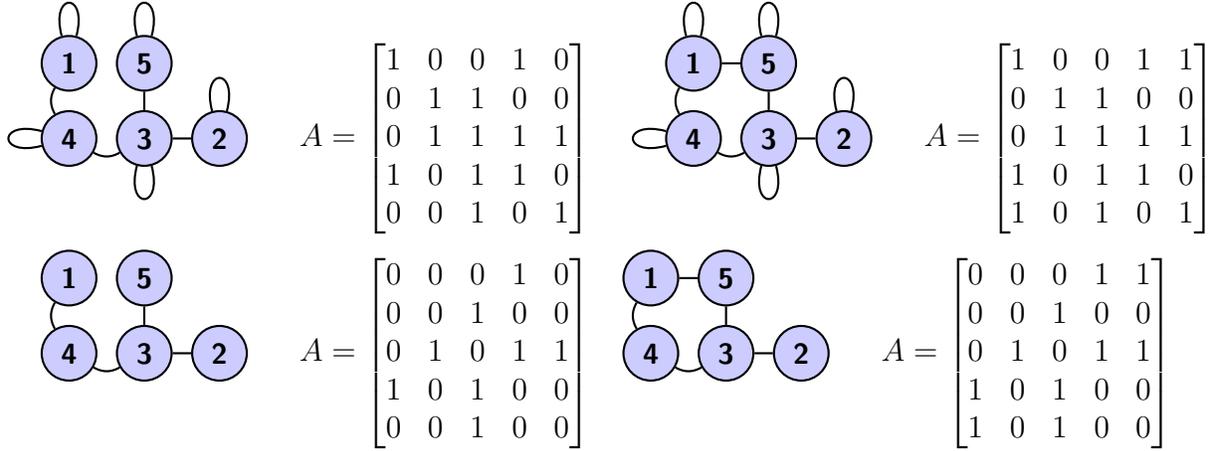
\begin{figure}[h!]
\begin{center}
\begin{tikzpicture}[ >=stealth',shorten >=0pt,auto,node distance=1.0cm,
  thick,main node/.style={circle,fill=blue!20,draw,font=\sffamily\small\bfseries}]
\tikzset{every loop/.style={}}

  \node[main node] (5){5};
  \node[main node](3) [below of=5]{3};
  \node[main node] (2) [right of=3] {2};
  \node[main node](4) [left of=3]{4};
  \node[main node](1) [above of=4]{1};
 \node[](6) at(4,-1) {$A=\begin{bmatrix}
1 & 0 & 0 & 1 & 0\\
0 & 1 & 1 & 0 & 0\\
0 & 1 & 1 & 1 & 1\\
1 & 0 & 1 & 1 & 0\\
0 & 0 & 1 & 0 & 1
\end{bmatrix}$};

  \path[every node/.style={font=\sffamily\small}]
           (1) edge [loop above] node {} (1)
           (2) edge [loop above] node {} (2)
           (3) edge [loop below] node {} (3)
           (4) edge [loop left] node {} (4)
           (5) edge [loop above] node {} (5)
    (5) edge node [right] {} (3)
       (3) edge [right] node[left] {} (2)
       (4) edge [bend right] node[left] {} (3)
        (4) edge [bend left] node[left] {} (1);
\end{tikzpicture}
\begin{tikzpicture}[ >=stealth',shorten >=0pt,auto,node distance=1.0cm,
  thick,main node/.style={circle,fill=blue!20,draw,font=\sffamily\small\bfseries}]
\tikzset{every loop/.style={}}

  \node[main node] (5){5};
  \node[main node](3) [below of=5]{3};
  \node[main node] (2) [right of=3] {2};
  \node[main node](4) [left of=3]{4};
  \node[main node](1) [above of=4]{1};
 \node[](6) at(4,-1) {$A=\begin{bmatrix}
1 & 0 & 0 & 1 & 1\\
0 & 1 & 1 & 0 & 0\\
0 & 1 & 1 & 1 & 1\\
1 & 0 & 1 & 1 & 0\\
1 & 0 & 1 & 0 & 1
\end{bmatrix}$};

  \path[every node/.style={font=\sffamily\small}]
           (1) edge [loop above] node {} (1)
           (2) edge [loop above] node {} (2)
           (3) edge [loop below] node {} (3)
           (4) edge [loop left] node {} (4)
           (5) edge [loop above] node {} (5)
    (5) edge node [right] {} (3)
    (5) edge node [right] {} (1)
       (3) edge [right] node[left] {} (2)
       (4) edge [bend right] node[left] {} (3)
        (4) edge [bend left] node[left] {} (1);
\end{tikzpicture}
\begin{tikzpicture}[ >=stealth',shorten >=0pt,auto,node distance=1.0cm,
  thick,main node/.style={circle,fill=blue!20,draw,font=\sffamily\small\bfseries}]
\tikzset{every loop/.style={}}

  \node[main node] (5){5};
  \node[main node](3) [below of=5]{3};
  \node[main node] (2) [right of=3] {2};
  \node[main node](4) [left of=3]{4};
  \node[main node](1) [above of=4]{1};
 \node[](6) at(4,-1) {$A=\begin{bmatrix}
 0 & 0 & 0 & 1 & 0\\
  0 & 0 & 1 & 0 & 0\\
 0 & 1 & 0 & 1 & 1\\
 1 & 0 & 1 & 0 & 0\\
 0 & 0 & 1 & 0 & 0
 \end{bmatrix}$};

  \path[every node/.style={font=\sffamily\small}]
    (5) edge node [right] {} (3)
       (3) edge [right] node[left] {} (2)
       (4) edge [bend right] node[left] {} (3)
        (4) edge [bend left] node[left] {} (1);
\end{tikzpicture}
\begin{tikzpicture}[ >=stealth',shorten >=0pt,auto,node distance=1.0cm,
  thick,main node/.style={circle,fill=blue!20,draw,font=\sffamily\small\bfseries}]
\tikzset{every loop/.style={}}

  \node[main node] (5){5};
  \node[main node](3) [below of=5]{3};
  \node[main node] (2) [right of=3] {2};
  \node[main node](4) [left of=3]{4};
  \node[main node](1) [above of=4]{1};
 \node[](6) at(4,-1) {$A=\begin{bmatrix}
0 & 0 & 0 & 1 & 1\\
0 & 0 & 1 & 0 & 0\\
0 & 1 & 0 & 1 & 1\\
1 & 0 & 1 & 0 & 0\\
1 & 0 & 1 & 0 & 0
\end{bmatrix}$};

  \path[every node/.style={font=\sffamily\small}]
    (5) edge node [right] {} (3)
    (5) edge node [right] {} (1)
       (3) edge [right] node[left] {} (2)
       (4) edge [bend right] node[left] {} (3)
        (4) edge [bend left] node[left] {} (1);
\end{tikzpicture}
\end{center} 
\caption{Graphs and adjacency matrices obtained as in \eqref{loopsnotransformunweighted} (top left),  \eqref{loopsabsunweighted} (top right),  \eqref{noloopsnotransformunweighted} (bottom left), and \eqref{noloopsabsunweighted} (bottom right). In all cases, $C$ is as in \eqref{eq:toyexample} and $\theta=0.25$.}
\label{fig2}
\end{figure}

The outcome $A$ depends, of course, not only on $C$ but also on $\theta$ which is indeed another parameter in our extensive tests. To illustrate this approach, we draw in Figure \ref{fig2} the graphs and adjacency matrices constructed for each of the four variants starting from \eqref{eq:toyexample} and for $\theta=0.25$. 


 \subsubsection{Weighted thresholded correlation matrix}
 
 The previous procedure can be simply modified to output a weighte graph. To this goal, we can simply keep the original weight on an edge if it is above the threshold $\theta$ (rather than mapping it to $1$), while still mapping the weight to $0$ (which is tantamount to removing the edge) when not.
 
 Mathematically, denoting by $\circ$ the Schur (elementwise) product of two matrices, this is realized as follows for each of the four possibilities described above:
  \begin{eqnarray}
 A &=& [C > \theta] \circ C, \label{loopsnotransformweighted}\\
 A &=& [|C| > \theta ] \circ |C|, \label{loopsabsweighted}\\
 A &=& [(C - I) > \theta] \circ (C-I), \label{noloopsnotransformweighted}\\ 
 A &=& [(|C|-I) > \theta] \circ (|C|-I). \label{noloopsabsweighted}
 \end{eqnarray}
The output is now an adjacency matrix of a weighted graph.

Again starting from \eqref{eq:toyexample} and with $\theta=0.25$, the outputs in terms of adjacency matrices are respectively
$$ A=\begin{bmatrix}
 1  & 0 &   0 &   0.4683  & 0\\
   0 &   1 &   0.4373  &  0 &   0\\
   0  &  0.4373 &   1 &   0.4245 &   0.4108\\
    0.4683  & 0  &  0.4245   & 1  & 0\\
   0 &  0  &  0.4108  & 0  &  1
\end{bmatrix}, \qquad A = \begin{bmatrix}
 1  & 0 &   0 &   0.4683  & 0.2583\\
   0 &   1 &   0.4373  &  0 &   0\\
   0  &  0.4373 &   1 &   0.4245 &   0.4108\\
    0.4683  & 0  &  0.4245   & 1  & 0\\
   0.2583 &  0  &  0.4108  & 0  &  1
\end{bmatrix},$$
$$ A=\begin{bmatrix}
 0  & 0 &   0 &   0.4683  & 0\\
   0 &   0 &   0.4373  &  0 &   0\\
   0  &  0.4373 &  0 &   0.4245 &   0.4108\\
    0.4683  & 0  &  0.4245   & 0  & 0\\
   0 &  0  &  0.4108  & 0  &  0
\end{bmatrix}, \qquad A = \begin{bmatrix}
 0  & 0 &   0 &   0.4683  & 0.2583\\
   0 &   0 &   0.4373  &  0 &   0\\
   0  &  0.4373 &   0 &   0.4245 &   0.4108\\
    0.4683  & 0  &  0.4245   & 0  & 0\\
   0.2583 &  0  &  0.4108  & 0  & 0
\end{bmatrix}.$$

We do not draw the graphs in this case as they are precisely the same as in Figure \ref{fig2}, except that now each edge or loop carries a possibly different weight (not represented in Figure \ref{fig2}).

\subsubsection{Minimal spanning tree}

For this method, we filter important information by constructing a minimal spanning tree (MST), starting from the correlation matrix. MSTs are a class of graphs that connect all vertices by placing an edge among the most correlated pairs without forming any cycles. MSTs tend to retain only significant correlations. Analyzing the tree structure, as a representation of the market can provide insights into the stability and state of the market and predict how shocks will propagate through a network. We use Prim's algorithm for computing the MST from the correlation matrix, as suggested in \cite{HGW09} after a comparison with other equivalent algorithms.

\subsubsection{Other options}
There are, of course, many other potential approaches that we did not include in our experiments. In particular, the Planar Maximal Filtering Graph (PMFG) has been advocated in the previous literature \cite{TMAM07}. For this reason, we have tested it. We have found that it is an extremely slow preprocessing step, and therefore, it does not fit with our criteria as it fails the Computational efficiency requirement. However, because it had been used previously, we have run it for a selected subsets of centrality measures. As we will see in more detail below, the PMFG is never appearing among the best 100 methods, neither when comparing methods based on cumulative return nor when ranking them based on Sharpe Ratios. We therefore believe that it does not bring any particularly precious added value to the idea of centrality measure based portfolio selection.

\subsection{Determining the threshold parameter $\theta$}

When building the adjacency matrix $A$ with a threshold-based filtering, the value of the threshold $\theta$ obviously play a crucial role. There is a tension here between removing some of the edges, thus somehow filtering noise and retaining relevant information only, and not losing too much information coming from the original (unfiltered) correlation matrix $C$.

We have tried the values $\theta = \frac{k}{10}$ for $k=0,1,2,\dots,9$. Heuristically, as we will discuss below, values of $\theta$ around $0.5$ seem to achieve the best performances.

\subsection{Determining the centrality parameter $\alpha$}

Several of the centrality measures that we employ, e.g. Katz or NBTW, depend on a parameter $\alpha$ that must lie in an interval of the form $(0,\alpha_{\max})$. For example, for classical Katz centrality, $\alpha_{\max} = \rho(A)$ is the largest eigenvalue of the adjacency matrix $A$, while for NBTW $\alpha_{\max}$ is the smallest eigenvalue of a certain matrix polynomial associated with the graph \cite{GHN18}. For exponential centrality we set $\alpha_{\max}=1$.

To test the behaviour of these centrality measures, and the stability of their performances with respect to variation in the choice of $\alpha$, we have followed the procedure below:
\begin{enumerate}
\item We computed $\alpha_{\max}$;
\item We tested the centrality for the values $\alpha = \frac{k}{10} \alpha_{\max}$ for $k=1,2,\dots,9$.
\end{enumerate}
For Katz centrality, we also tested the specially predetermined value $$\alpha = \frac{1 - \exp(-\alpha_{\max})}{\alpha_{\max}},$$ which has the property of minimizing the distance between the rankings generated by Katz and exponential centrality; we will refer to this particular version of Katz as ``Katz-min''.  See \cite{AHH16} for more details on this approach to choose $\alpha$.

\section{Data analysis}
For the empirical application we focus on the S\&P 500 index and its constituents. The $S\&P 500$ index consists of the $500$ companies with the largest capitalisation across multiple industry sectors. We consider an investor who invest in a maximum of $m$ stocks from the constituent list of the S\&P 500. The data employed cover the period from January 1990 to December 2017. We use the data in a daily frequency. The train set is equal to $1990:2016$ and the test set is $1991:2017$. For each year $t$ in the train set period the following procedure is followed.
\begin{enumerate}
	\item Using the returns of a specific year, say $t=1990$, we calculate the centrality measures described in Section $2.1$ for all stocks of S\&P  500 in the end of the year.	
	\item We sort the stocks based on their centrality measure and we select (a) the $m$ stocks with the highest centrality (central nodes in the graph) and (b) the $m$ stocks with the lowest centrality (peripheral nodes in the graph). 
	\item We calculate the weights of equally-weighted and optimal portfolios for these $m$ stocks based on their returns in year$t$.
	\item Using the weights calculated in step 3, we invest in these portfolios in year $t+1$, and calculate the portfolios realised returns.  
\end{enumerate}

We construct the following portfolios: equal weights (EW), minimum variance (Min Var) and mean-variance (Mean-Var). In the case of the optimal portfolios (Min Var and Mean-Var) we consider the case of long positions only (LO) and the case of both long and short positions (LS). The weights are constrained to be lower or equal to $0.25$ in the (LO) case and between $-0.25$ and $0.25$ in the (LS) case.

We repeat the procedure described above and we and end up having portfolios of stocks employing their daily returns for a period of $27$ years, which is quite representative as it covers recessions, expansions and also periods of high market volatility. We consider portfolios of $m=10$ stocks and use as a benchmark for evaluating the proposed portfolios we use the S\&P 500 index.
  
In addition, we employ the following values regarding the parameters of the graphs:
\begin{itemize}
    \item Correlation matrix tansformation: (A1) no transformation, (A2) 
    positive part of the correlation matrix defined as by $C^{+}_{i,j} = \max \{ C_{i,j}, 0 \}$ (subplus function of the correlation matrix), (A3) positive part of the negative elements of the correlation matrix defined as by $-C^{+}_{i,j} = \max \{ -C_{i,j},0 \}$ (subplus function of the negative elements of the correlation matrix), (A4) absolute value of correlation matrix.
    \item Centrality  measure: degree, Katz, subgraph, exponential, exponential subgraph, NBTW, NBTW subgraph, NBTW exponential, NBTW exponential subgraph and betweenness centrality.
    \item Threshold of correlation matrix $\theta$: $0.0, 0.1, \ldots, 0.9$. 
    \item Adjacency matrix options: \\ 
      \begin{eqnarray*}
     (1) \  A &=& C > \theta, \\ 
     (2) \  A &=& |C| > \theta, \\ 
     (3) \  A &=& (C - I) > \theta, \\ 
     (4) \  A &=& (|C|-I) > \theta. \\ 
     (5) \  A &=& [C > \theta] \circ C, \\ 
     (6) \  A &=& [|C| > \theta ] \circ |C|, \\ 
     (7) \  A &=& [(C - I) > \theta] \circ (C-I), \\ 
     (8) \  A &=& [(|C|-I) > \theta] \circ (|C|-I). \\ 
      \end{eqnarray*}
where $A$ is the adjacency matrix, $C$ is the correlation matrix and $\theta$ is the threshold employed. In this case an edge between two nodes exist if and only if the cross-correlation between two nodes is higher than the specified threshold value $\theta$. 
    \item $\alpha$ = $0.1 \cdot \alpha_{\max}, \ldots, 0.9 \cdot \alpha_{\max}$. 
\end{itemize}

The portfolios are evaluated using the performance evaluation criteria described in Section 2.3 and more importantly using Sharpe Ratio and expected return.

In Tables $1$ to $9$ we present the best portfolios based on the Sharpe Ratio (SR) and the Expected Return (ER) using different centrality measures, adjacency matrix options, and other parameters; the parameter $\alpha$ is reported as a fraction of $\alpha_{\max}$. We also compute a range of portfolio metrics in order to facilitate their comparison. These metrics include expected return,  standard deviation, Sharpe Ratio, Value at Risk (VaR), Conditional Value at Risk (CVaR), Maximum Drawdown (MaxDDar), Omega Ratio, Sortino Ratio and Upside Potential (UP).

Table 1 showcases the top thirty portfolios ranked by their Sharpe Ratio (SR), ranging from $0.7833$ to $0.8295$. Notably, all these portfolios are constructed using peripheral stocks. Centrality measures such as Exponential, Katz Min, Exponential Subgraph, and NBTW Subgraph exhibit the highest SRs. Equally weighted portfolios and minimum variance strategies stand out for their strong performance, often out-performing mean-variance portfolios. The threshold values commonly used are $0.5$ and $0.6$ and among the adjacency matrix options, the configurations (7) (no loops, no transform and weighted graphs), (3) (no loops, no transform and unweighted graphs) and (4) (no loops, absolute value transform and unweighted graphs), have the highest frequency. In approximately $56\%$ of the cases the shrinkage is employed.

Table 2 depicts the top twenty portfolios ranked by their annualised  expected return (ER) for the A1 cases (no transform of the correlation matrix). Expected returns range from $0.18$ to $0.20$. Remarkably, all portfolios, except one, consist of central stocks, with only a single portfolio incorporating peripheral stocks. In this category, subgraph centralities notably outperform other centrality measures. Specifically, Exponential Subgraph, NBTW Subgraph, and Katz Subgraph are prominent among the top 20 portfolios. It is noteworthy that mean-variance portfolios are employed in most of the cases, with equally weighted portfolios appearing only once. The threshold value commonly utilized in the majority of cases is $0.7$, accompanied by low values of the parameter $\alpha$. Shrinkage is not utilized in any of the cases. Among the adjacency matrix options, configurations (2) (loops, absolute value transform, and unweighted graphs) and (4) (no loops, absolute value transform, and unweighted graphs) exhibit the highest frequency.

In Table 3 we see  the top twenty portfolios ranked by their Sharpe Ratio (SR), for the A1 cases (no initial transform of the correlation matrix). Sharpe Ratios are ranging from $0.74$ to $0.83$. Notably, all these portfolios are constructed using peripheral stocks. Centrality measures such as Katz Min, NBTW Subgraph and Exponential Subgraph exhibit the highest SRs. We see that equally weighted portfolios, mean-variance and minimum variance portfolios to stand out for their strong performance. The threshold values commonly used are $0.5$ and $0.6$, and in all cases shrinkage is not used. Among the adjacency matrix options, the configurations the configurations (4) (no loops, absolute value transform and unweighted graphs) and (2) (loops, absolute value transform, and unweighted graphs) have the highest frequency. 

In Tables 4, 6 and 8 we see a similar picture as in Table 2.
Table 4 depicts the top twenty portfolios ranked by their annualised  expected return (ER) for the A2 cases (positive part of the correlation matrix), Table 6 for the A3 cases (subplus function of the negative elements of the correlation matrix) and Table 8 for the 
A4 (absolute value of correlation matrix). Expected returns range from $0.18$ to $0.20$. Again, all portfolios consist mainly of the most central stocks, with only a few exceptions of portfolios incorporating peripheral stocks. In these categories, subgraph centralities notably outperform other centrality measures. Specifically, Exponential Subgraph, NBTW Subgraph, and Katz Subgraph are prominent among the top twenty portfolios. In some exceptions we see also Exponential and Katz centrality. As before mean-variance portfolios are employed in most of the cases, with equally weighted portfolios appearing only once. The threshold value commonly utilized in the majority of cases is $0.7$.  Shrinkage is utilized in all cases in Table 4 with A2 and Table 8 with A4, but not in any of the cases in Table 6 with A3. Among the adjacency matrix options, in Table 4 configurations (2) (loops, absolute value transform, and unweighted graphs) and (4) (no loops, absolute value transform, and unweighted graphs) exhibit the highest frequency and in Tables 6 and 8 configurations (1) (loops, no transform, unweighted) and (3) (no loops, no transform and unweighted graphs).

In Tables 5, 7 and 9 we see a similar picture as in Table 3. Table 5 depicts the top twenty portfolios ranked by their annualised  Sharpe Ratios for the A2 cases (positive part of the correlation matrix), Table 7 for the A3 cases (subplus function of the negative elements of the correlation matrix) and Table 8 for the 
A4 (absolute value of correlation matrix). Sharpe Ratios returns range from $0.76$ to $0.83$. Again as in Table 1 and Table 3, all portfolios consist mainly of the most peripheral stocks. Centrality measures such as Exponential, Katz Min, Exponential Subgraph, and NBTW Subgraph exhibit the highest SRs. In addition, the equally weighted portfolios and minimum variance strategies again appear to have the higher frequency among the most strong portfolios with mean-variance portfolios appearing less frequently. The threshold values commonly used are between $0.4$ and $0.8$ with $0.5$ and $0.6$ having the higher frequency. Regarding the adjacency matrix options, the configurations (4) and (8) appear mostly in Table 5, configurations (1), (3), (5) and (7) in Table 7 and Table 9.
Shrinkage is utilized in all cases in Table 5 and 9, but not in any of the cases in Table 7.

\begin{sidewaystable}[h!]
\caption{Best 30 methods for SR}  
\bigskip
\centering 
\resizebox{\linewidth}{!}{
\begin{tabular}{llllllrrrrrrrrrrrl}
  \hline
Portfolio & Centrality & AM & C/P & TV & alpha & CR & ER & SD & {\bf SR} & VaR005 & CVaR005 & MaxDDar & Omega & Sortino & UP & Graph & Shrink \\ 
\hline
EW & Exponential & \eqref{noloopsnotransformweighted} & peripheral & 0.5 & 0.9 & 4.91 & 0.18 & 0.18 & {\bf 0.8295} & -0.27 & -0.40 & 79.49 & 1.16 & 0.08 & 0.55 & ANV & Y \\ 
  Min Var (LS) & Katz Min & \eqref{noloopsabsunweighted} & peripheral & 0.5 & NaN & 4.75 & 0.17 & 0.17 & {\bf 0.8275} & -0.26 & -0.39 & 30.45 & 1.16 & 0.08 & 0.56 & ANV & N \\ 
  Min Var (LS) & Katz Min & \eqref{noloopsnotransformunweighted} & peripheral & 0.5 & NaN & 4.75 & 0.17 & 0.17 & {\bf 0.8275} & -0.26 & -0.39 & 30.45 & 1.16 & 0.08 & 0.56 & ANV & N \\ 
  EW & Exponential Subgraph & \eqref{noloopsnotransformweighted} & peripheral & 0.5 & 1 & 5.05 & 0.18 & 0.19 & {\bf 0.8225} & -0.28 & -0.42 & 64.31 & 1.16 & 0.08 & 0.55 & ANV & Y \\ 
  EW & Exponential & \eqref{loopsnotransformweighted} & peripheral & 0.5 & 1 & 5.20 & 0.19 & 0.20 & {\bf 0.8167} & -0.29 & -0.43 & 70.59 & 1.16 & 0.08 & 0.55 & ANV & Y \\ 
  EW & Exponential Subgraph & \eqref{noloopsnotransformunweighted} & peripheral & 0.5 & 0.5 & 4.94 & 0.18 & 0.19 & {\bf 0.8098} & -0.28 & -0.42 & 75.46 & 1.15 & 0.07 & 0.55 & ANV & Y \\ 
  EW & Exponential & \eqref{noloopsnotransformunweighted} & peripheral & 0.5 & 0.5 & 4.99 & 0.18 & 0.19 & {\bf 0.8088} & -0.28 & -0.41 & 59.33 & 1.16 & 0.07 & 0.56 & ANV & Y \\ 
    Min Var (LS) & Exponential & \eqref{noloopsabsweighted} & peripheral & 0.7 & 0.2 & 4.50 & 0.16 & 0.17 & {\bf 0.8051} & -0.24 & -0.38 & 43.10 & 1.16 & 0.07 & 0.54 & ANV & Y \\ 
  Mean-Var (LO) & Exponential & \eqref{noloopsabsunweighted} & peripheral & 0.6 & 0.5 & 4.35 & 0.16 & 0.16 & {\bf 0.8044} & -0.25 & -0.35 & 38.63 & 1.15 & 0.07 & 0.57 & ANV & Y \\ 
    EW & NBTW Subgraph & \eqref{noloopsnotransformunweighted} & peripheral & 0.5 & 0.3 & 5.01 & 0.18 & 0.19 & {\bf 0.8021} & -0.29 & -0.43 & 48.48 & 1.16 & 0.07 & 0.56 & ANV & N \\ 
      EW & NBTW Subgraph & \eqref{noloopsabsunweighted} & peripheral & 0.5 & 0.3 & 5.01 & 0.18 & 0.19 & {\bf 0.8021} & -0.29 & -0.43 & 48.48 & 1.16 & 0.07 & 0.56 & ANV & N \\       
  Min Var (LO) & Exponential & \eqref{noloopsabsunweighted} & peripheral & 0.6 & 0.5 & 4.07 & 0.15 & 0.15 & {\bf 0.8013} & -0.22 & -0.34 & 36.80 & 1.15 & 0.07 & 0.56 & ANV & Y \\ 
      EW & NBTW Subgraph & \eqref{noloopsnotransformunweighted} & peripheral & 0.4 & 0.3 & 5.02 & 0.18 & 0.19 & {\bf 0.7982} & -0.28 & -0.44 & 72.52 & 1.16 & 0.07 & 0.54 & ANV & N \\   
  EW & NBTW Subgraph & \eqref{noloopsabsunweighted} & peripheral & 0.4 & 0.3 & 5.02 & 0.18 & 0.19 & {\bf 0.7982} & -0.28 & -0.44 & 72.52 & 1.16 & 0.07 & 0.54 & ANV & N\\ 
    EW & Exponential & \eqref{noloopsabsweighted} & peripheral & 0.5 & 0.9 & 5.00 & 0.18 & 0.19 & {\bf 0.7970} & -0.28 & -0.42 & 60.74 & 1.16 & 0.07 & 0.55 & ANV & Y \\ 
  EW & Exponential Subgraph & \eqref{noloopsabsweighted} & peripheral & 0.5 & 0.7 & 5.21 & 0.19 & 0.20 & {\bf 0.7966} & -0.28 & -0.45 & 86.76 & 1.16 & 0.07 & 0.54 & ANV & N \\           
  EW & Exponential Subgraph & \eqref{noloopsnotransformweighted} & peripheral & 0.5 & 0.7 & 5.21 & 0.19 & 0.20 & {\bf 0.7966} & -0.28 & -0.45 & 86.76 & 1.16 & 0.07 & 0.54 & ANV & N \\ 
  Min Var (LS) & Exponential & \eqref{noloopsabsunweighted} & peripheral & 0.6 & 0.5 & 4.07 & 0.15 & 0.15 & {\bf 0.7942} & -0.23 & -0.34 & 34.45 & 1.15 & 0.07 & 0.56 & ANV & Y \\ 
  Min Var (LO) & Exponential Subgraph & \eqref{noloopsabsunweighted} & peripheral & 0.6 & 0.1 & 4.15 & 0.15 & 0.15 & {\bf 0.7913} & -0.23 & -0.35 & 56.09 & 1.15 & 0.07 & 0.56 & ANV & Y \\   
  EW & Exponential Subgraph & \eqref{noloopsnotransformweighted} & peripheral & 0.4 & 0.8 & 5.01 & 0.18 & 0.19 & {\bf 0.7909} & -0.28 & -0.42 & 74.12 & 1.16 & 0.07 & 0.54 & ANV & Y \\ 
  EW & Exponential Subgraph & \eqref{noloopsnotransformweighted} & peripheral & 0.5 & 0.5 & 4.91 & 0.18 & 0.19 & {\bf 0.7901} & -0.28 & -0.42 & 68.83 & 1.15 & 0.07 & 0.55 & ANV & Y \\ 
  EW & Exponential Subgraph & \eqref{noloopsnotransformweighted} & peripheral & 0.5 & 0.7 & 4.75 & 0.17 & 0.18 & {\bf 0.7872} & -0.28 & -0.41 & 67.95 & 1.15 & 0.07 & 0.56 & ANV & Y \\ 
      Mean-Var (LO) & Exponential Subgraph & \eqref{loopsabsunweighted} & peripheral & 0.6 & 0.9 & 4.48 & 0.16 & 0.17 & {\bf 0.7863} & -0.25 & -0.38 & 40.07 & 1.15 & 0.07 & 0.56 & ANV & Y \\  
  Mean-Var (LS) & Exponential Subgraph & \eqref{loopsnotransformunweighted} & peripheral & 0.6 & 0.1 & 4.69 & 0.17 & 0.18 & {\bf 0.7861} & -0.27 & -0.41 & 47.54 & 1.15 & 0.07 & 0.56 & ANV & N \\ 
  Mean-Var (LS) & Exponential Subgraph & \eqref{loopsabsunweighted} & peripheral & 0.6 & 0.1 & 4.69 & 0.17 & 0.18 & {\bf 0.7861} & -0.27 & -0.41 & 47.54 & 1.15 & 0.07 & 0.56 & ANV & N \\ 
    EW & Exponential Subgraph & \eqref{loopsnotransformunweighted} & peripheral & 0.5 & 1 & 4.62 & 0.16 & 0.18 & {\bf 0.7859} & -0.26 & -0.40 & 50.94 & 1.15 & 0.07 & 0.55 & ANV & Y \\
  Min Var (LO) & Exponential Subgraph & \eqref{loopsabsunweighted} & peripheral & 0.6 & 0.1 & 4.22 & 0.15 & 0.16 & {\bf 0.7852} & -0.25 & -0.36 & 36.35 & 1.15 & 0.07 & 0.56 & ANV & N \\   
  Min Var (LO) & Exponential Subgraph & \eqref{loopsnotransformunweighted} & peripheral & 0.6 & 0.1 & 4.22 & 0.15 & 0.16 & {\bf 0.7852} & -0.25 & -0.36 & 36.35 & 1.15 & 0.07 & 0.56 & ANV & N \\   
    Min Var (LO) & Exponential & \eqref{noloopsabsweighted} & peripheral & 0.7 & 0.2 & 4.40 & 0.16 & 0.17 & {\bf 0.7844} & -0.25 & -0.39 & 43.03 & 1.15 & 0.07 & 0.54 & ANV & Y \\ 
  EW & Exponential Subgraph & \eqref{loopsabsweighted} & peripheral & 0.5 & 0.6 & 4.96 & 0.18 & 0.19 & {\bf 0.7833} & -0.28 & -0.44 & 74.48 & 1.16 & 0.07 & 0.54 & ANV & N\\ 
\hline
\end{tabular}
}
\end{sidewaystable}

\begin{sidewaystable}[h!]
\caption{ER 20 - A1}  
\bigskip
\centering 
\resizebox{\linewidth}{!}{
\begin{tabular}{llllllrrrrrrrrrrrl}
  \hline
Portfolio & Centrality & AM & C/P & TV & alpha & CR & ER & SD & SR & VaR005 & CVaR005 & MaxDDar & Omega & Sortino & UP & Graph & Shrink\\ 
  \hline
Mean-Var (LS) & Exponential Subgraph & \eqref{loopsabsunweighted} & central & 0.7 & 0.1 & 5.51 & 0.20 & 0.28 & 0.61 & -0.39 & -0.63 & 98.13 & 1.12 & 0.06 & 0.52 & ANV & N \\ 
  Mean-Var (LS) & NBTW Subgraph & \eqref{noloopsabsunweighted} & central & 0.7 & 0.9 & 5.42 & 0.19 & 0.27 & 0.61 & -0.38 & -0.62 & 98.13 & 1.13 & 0.06 & 0.51 & ANV & N \\ 
  Mean-Var (LS) & NBTW Subgraph & \eqref{noloopsabsunweighted} & central & 0.7 & 0.4 & 5.38 & 0.19 & 0.27 & 0.61 & -0.38 & -0.62 & 98.13 & 1.12 & 0.06 & 0.51 & ANV & N \\ 
  Mean-Var (LS) & Exponential Subgraph & \eqref{noloopsabsunweighted} & central & 0.7 & 0.1 & 5.33 & 0.19 & 0.28 & 0.59 & -0.39 & -0.63 & 98.13 & 1.12 & 0.06 & 0.51 & ANV & N\\ 
  Mean-Var (LS) & Katz Subgraph & \eqref{loopsabsunweighted} & central & 0.7 & 0.8 & 5.30 & 0.19 & 0.28 & 0.58 & -0.41 & -0.64 & 98.13 & 1.12 & 0.05 & 0.51 & ANV & N \\ 
  Mean-Var (LS) & NBTW Subgraph & \eqref{loopsabsunweighted} & central & 0.7 & 0.6 & 5.30 & 0.19 & 0.28 & 0.58 & -0.41 & -0.64 & 98.13 & 1.12 & 0.05 & 0.52 & ANV & N \\ 
  Mean-Var (LS) & Katz Subgraph & \eqref{loopsabsunweighted} & central & 0.7 & 0.7 & 5.27 & 0.19 & 0.28 & 0.58 & -0.40 & -0.64 & 98.13 & 1.12 & 0.05 & 0.51 & ANV & N \\ 
  Mean-Var (LS) & Exponential Subgraph & \eqref{noloopsabsunweighted} & central & 0.7 & 0.3 & 5.26 & 0.19 & 0.28 & 0.58 & -0.40 & -0.63 & 94.28 & 1.12 & 0.05 & 0.51 & ANV & N \\ 
  Mean-Var (LS) & NBTW Subgraph & \eqref{noloopsabsunweighted} & central & 0.7 & 0.5 & 5.26 & 0.19 & 0.27 & 0.59 & -0.38 & -0.62 & 98.13 & 1.12 & 0.06 & 0.51 & ANV & N \\ 
  Mean-Var (LS) & Exponential Subgraph & \eqref{loopsabsunweighted} & central & 0.7 & 0.2 & 5.25 & 0.19 & 0.28 & 0.58 & -0.40 & -0.62 & 94.28 & 1.12 & 0.05 & 0.52 & ANV  & N\\ 
  Mean-Var (LS) & Katz Subgraph & \eqref{noloopsabsunweighted} & central & 0.7 & 0.8 & 5.25 & 0.19 & 0.28 & 0.57 & -0.40 & -0.64 & 98.13 & 1.12 & 0.05 & 0.51 & ANV & N \\ 
  Mean-Var (LS) & NBTW Subgraph & \eqref{noloopsabsunweighted} & central & 0.7 & 0.3 & 5.24 & 0.19 & 0.27 & 0.59 & -0.38 & -0.62 & 98.13 & 1.12 & 0.06 & 0.51 & ANV & N \\ 
  Mean-Var (LS) & NBTW Subgraph & \eqref{loopsabsunweighted} & central & 0.7 & 0.7 & 5.24 & 0.19 & 0.28 & 0.57 & -0.41 & -0.64 & 98.13 & 1.11 & 0.05 & 0.51 & ANV & N \\ 
  Mean-Var (LS) & NBTW Subgraph & \eqref{noloopsabsunweighted} & central & 0.7 & 0.1 & 5.23 & 0.19 & 0.27 & 0.59 & -0.38 & -0.62 & 98.13 & 1.12 & 0.06 & 0.51 & ANV & N \\ 
  Mean-Var (LS) & Katz Subgraph & \eqref{noloopsabsunweighted} & central & 0.7 & 0.7 & 5.21 & 0.19 & 0.28 & 0.57 & -0.41 & -0.64 & 98.13 & 1.11 & 0.05 & 0.51 & ANV & N \\ 
  EW & Exponential Subgraph & \eqref{noloopsabsweighted} & peripheral & 0.5 & 0.7 & 5.21 & 0.19 & 0.20 & 0.80 & -0.28 & -0.45 & 86.76 & 1.16 & 0.07 & 0.54 & ANV & N \\ 
  Mean-Var (LS) & NBTW Subgraph & \eqref{noloopsabsunweighted} & central & 0.7 & 0.2 & 5.21 & 0.19 & 0.28 & 0.58 & -0.38 & -0.62 & 98.13 & 1.12 & 0.06 & 0.51 & ANV & N \\ 
  Mean-Var (LS) & Exponential Subgraph & \eqref{loopsabsunweighted} & central & 0.7 & 0.3 & 5.19 & 0.19 & 0.28 & 0.57 & -0.40 & -0.63 & 94.28 & 1.11 & 0.05 & 0.52 & ANV & N \\ 
  Min Var (LS) & NBTW Subgraph & \eqref{noloopsabsunweighted} & central & 0.7 & 0.4 & 5.17 & 0.18 & 0.28 & 0.56 & -0.37 & -0.64 & 66.09 & 1.12 & 0.05 & 0.48 & ANV & N \\ 
  Mean-Var (LS) & NBTW Subgraph & \eqref{loopsabsunweighted} & central & 0.7 & 0.9 & 5.17 & 0.18 & 0.28 & 0.56 & -0.41 & -0.64 & 98.13 & 1.11 & 0.05 & 0.51 & ANV & N \\ 
   \hline
\end{tabular}
}
\end{sidewaystable}

\begin{sidewaystable}[h!]
\caption{SR 20 - A1} 
\bigskip
\centering 
\resizebox{\linewidth}{!}{
\begin{tabular}{llllllrrrrrrrrrrrl}
  \hline
Portfolio & Centrality & AM & C/P & TV & alpha & CR & ER & SD & SR & VaR005 & CVaR005 & MaxDDar & Omega & Sortino & UP & Graph & Shrink \\ 
\hline
Min Var (LS) & Katz Min & \eqref{noloopsabsunweighted} & peripheral & 0.5 & NaN & 4.75 & 0.17 & 0.17 & 0.83 & -0.26 & -0.39 & 30.45 & 1.16 & 0.08 & 0.56 & ANV & N \\ 
  EW & NBTW Subgraph & \eqref{noloopsabsunweighted} & peripheral & 0.5 & 0.3 & 5.01 & 0.18 & 0.19 & 0.80 & -0.29 & -0.43 & 48.48 & 1.16 & 0.07 & 0.56 & ANV & N \\ 
  EW & NBTW Subgraph & \eqref{noloopsabsunweighted} & peripheral & 0.4 & 0.3 & 5.02 & 0.18 & 0.19 & 0.80 & -0.28 & -0.44 & 72.52 & 1.16 & 0.07 & 0.54 & ANV & N\\ 
  EW & Exponential Subgraph & \eqref{noloopsabsweighted} & peripheral & 0.5 & 0.7 & 5.21 & 0.19 & 0.20 & 0.80 & -0.28 & -0.45 & 86.76 & 1.16 & 0.07 & 0.54 & ANV & N \\ 
  Mean-Var (LS) & Exponential Subgraph & \eqref{loopsabsunweighted} & peripheral & 0.6 & 0.1 & 4.69 & 0.17 & 0.18 & 0.79 & -0.27 & -0.41 & 47.54 & 1.15 & 0.07 & 0.56 & ANV & N \\ 
  Min Var (LO) & Exponential Subgraph & \eqref{loopsabsunweighted} & peripheral & 0.6 & 0.1 & 4.22 & 0.15 & 0.16 & 0.78 & -0.25 & -0.36 & 36.35 & 1.15 & 0.07 & 0.56 & ANV & N \\ 
  EW & Exponential Subgraph & \eqref{loopsabsweighted} & peripheral & 0.5 & 0.6 & 4.96 & 0.18 & 0.19 & 0.78 & -0.28 & -0.44 & 74.48 & 1.16 & 0.07 & 0.54 & ANV & N\\ 
  Mean-Var (LO) & Exponential Subgraph & \eqref{loopsabsunweighted} & peripheral & 0.6 & 0.1 & 4.50 & 0.16 & 0.17 & 0.78 & -0.27 & -0.39 & 42.86 & 1.15 & 0.07 & 0.56 & ANV & N\\ 
  EW & Exponential Subgraph & \eqref{noloopsabsunweighted} & peripheral & 0.4 & 1 & 5.03 & 0.18 & 0.20 & 0.78 & -0.29 & -0.45 & 65.42 & 1.15 & 0.07 & 0.53 & ANV & N \\ 
  EW & Exponential Subgraph & \eqref{noloopsabsweighted} & peripheral & 0.6 & 0.5 & 4.72 & 0.17 & 0.18 & 0.77 & -0.28 & -0.41 & 42.29 & 1.15 & 0.07 & 0.56 & ANV & N\\ 
  Mean-Var (LS) & Exponential & \eqref{loopsabsunweighted} & peripheral & 0.6 & 0.1 & 4.48 & 0.16 & 0.17 & 0.77 & -0.26 & -0.39 & 62.95 & 1.15 & 0.07 & 0.56 & ANV& N \\ 
  EW & Exponential Subgraph & \eqref{loopsabsunweighted} & peripheral & 0.6 & 0.8 & 4.98 & 0.18 & 0.20 & 0.77 & -0.29 & -0.45 & 64.50 & 1.15 & 0.07 & 0.54 & ANV& N \\ 
  EW & NBTW Subgraph & \eqref{noloopsabsunweighted} & peripheral & 0.4 & 0.7 & 4.80 & 0.17 & 0.19 & 0.76 & -0.28 & -0.44 & 71.72 & 1.15 & 0.07 & 0.54 & ANV & N \\ 
  EW & Exponential Subgraph & \eqref{noloopsabsunweighted} & peripheral & 0.4 & 0.5 & 4.87 & 0.17 & 0.20 & 0.76 & -0.28 & -0.44 & 63.66 & 1.15 & 0.07 & 0.53 & ANV & N \\ 
  Min Var (LS) & Exponential Subgraph & \eqref{loopsabsweighted} & peripheral & 0.8 & 0.6 & 4.12 & 0.15 & 0.16 & 0.75 & -0.24 & -0.36 & 52.75 & 1.14 & 0.07 & 0.56 & ANV & N \\ 
  Min Var (LO) & Exponential Subgraph & \eqref{loopsabsweighted} & peripheral & 0.8 & 0.6 & 4.11 & 0.15 & 0.16 & 0.75 & -0.25 & -0.36 & 58.75 & 1.14 & 0.07 & 0.56 & ANV & N \\ 
  Min Var (LS) & Exponential Subgraph & \eqref{loopsabsunweighted} & peripheral & 0.6 & 0.1 & 4.06 & 0.14 & 0.16 & 0.75 & -0.25 & -0.36 & 38.06 & 1.14 & 0.07 & 0.56 & ANV  & N\\ 
  Min Var (LO) & Katz Min & \eqref{noloopsabsunweighted} & peripheral & 0.5 & NaN & 4.40 & 0.16 & 0.18 & 0.75 & -0.27 & -0.40 & 39.83 & 1.14 & 0.07 & 0.55 & ANV & N \\ 
  EW & Exponential Subgraph & \eqref{noloopsabsweighted} & peripheral & 0.5 & 0.3 & 4.44 & 0.16 & 0.18 & 0.75 & -0.26 & -0.40 & 41.84 & 1.14 & 0.07 & 0.55 & ANV & N \\ 
  EW & NBTW Subgraph & \eqref{loopsabsweighted} & peripheral & 0.5 & 0.2 & 4.64 & 0.17 & 0.19 & 0.74 & -0.28 & -0.43 & 64.35 & 1.14 & 0.07 & 0.55 & ANV & N \\ 
\hline
\end{tabular}
}
\end{sidewaystable}

\begin{sidewaystable}[h!]
\caption{ER 20 - A2} 
\bigskip
\centering 
\resizebox{\linewidth}{!}{
\begin{tabular}{llllllrrrrrrrrrrrl}
  \hline
Portfolio & Centrality & AM & C/P & TV & alpha & CR & ER & SD & SR & VaR005 & CVaR005 & MaxDDar & Omega & Sortino & UP & Graph & Shrink \\ 
\hline
Mean-Var (LS) & Katz Subgraph & \eqref{loopsabsunweighted} & central & 0.7 & 0.7 & 5.48 & 0.20 & 0.28 & 0.61 & -0.40 & -0.63 & 98.13 & 1.12 & 0.06 & 0.52 & ANV & Y \\ 
  Mean-Var (LS) & Exponential Subgraph & \eqref{loopsabsunweighted} & central & 0.7 & 0.1 & 5.46 & 0.20 & 0.28 & 0.61 & -0.40 & -0.62 & 96.30 & 1.12 & 0.06 & 0.52 & ANV & Y \\ 
  Mean-Var (LS) & Exponential Subgraph & \eqref{noloopsabsunweighted} & central & 0.7 & 0.1 & 5.41 & 0.19 & 0.28 & 0.61 & -0.40 & -0.62 & 96.30 & 1.12 & 0.06 & 0.52 & ANV & Y \\ 
  Mean-Var (LS) & Exponential & \eqref{loopsabsunweighted} & central & 0.7 & 0.2 & 5.34 & 0.19 & 0.27 & 0.60 & -0.39 & -0.61 & 87.03 & 1.12 & 0.06 & 0.52 & ANV & Y \\ 
  Mean-Var (LS) & Exponential Subgraph & \eqref{noloopsabsunweighted} & central & 0.7 & 0.2 & 5.33 & 0.19 & 0.27 & 0.60 & -0.39 & -0.61 & 87.03 & 1.12 & 0.06 & 0.52 & ANV & Y \\ 
  Mean-Var (LS) & Katz Subgraph & \eqref{noloopsabsunweighted} & central & 0.7 & 0.7 & 5.32 & 0.19 & 0.28 & 0.59 & -0.41 & -0.63 & 98.13 & 1.12 & 0.06 & 0.52 & ANV & Y \\ 
  Mean-Var (LS) & NBTW Subgraph & \eqref{loopsabsunweighted} & central & 0.7 & 0.6 & 5.32 & 0.19 & 0.28 & 0.59 & -0.41 & -0.63 & 98.13 & 1.12 & 0.06 & 0.52 & ANV & Y \\ 
  Mean-Var (LS) & Exponential Subgraph & \eqref{noloopsabsunweighted} & central & 0.7 & 0.3 & 5.30 & 0.19 & 0.27 & 0.60 & -0.39 & -0.61 & 87.03 & 1.12 & 0.06 & 0.52 & ANV & Y \\ 
  Mean-Var (LS) & Exponential Subgraph & \eqref{loopsabsunweighted} & central & 0.7 & 0.3 & 5.30 & 0.19 & 0.27 & 0.60 & -0.39 & -0.61 & 87.03 & 1.12 & 0.06 & 0.52 & ANV & Y \\ 
  Mean-Var (LS) & NBTW Subgraph & \eqref{loopsabsunweighted} & central & 0.7 & 0.7 & 5.30 & 0.19 & 0.28 & 0.59 & -0.41 & -0.63 & 98.13 & 1.12 & 0.06 & 0.52 & ANV & Y \\ 
  Mean-Var (LS) & Exponential & \eqref{noloopsabsunweighted} & central & 0.7 & 0.2 & 5.29 & 0.19 & 0.27 & 0.60 & -0.39 & -0.61 & 87.03 & 1.12 & 0.06 & 0.52 & ANV & Y \\ 
  Mean-Var (LS) & Katz Subgraph & \eqref{noloopsabsunweighted} & central & 0.7 & 0.8 & 5.28 & 0.19 & 0.28 & 0.58 & -0.40 & -0.62 & 96.06 & 1.12 & 0.06 & 0.52 & ANV & Y \\ 
  Mean-Var (LS) & Katz Subgraph & \eqref{loopsabsunweighted} & central & 0.7 & 0.6 & 5.26 & 0.19 & 0.28 & 0.58 & -0.40 & -0.62 & 98.13 & 1.12 & 0.06 & 0.52 & ANV & Y \\ 
  Mean-Var (LS) & Exponential Subgraph & \eqref{loopsabsunweighted} & central & 0.7 & 0.2 & 5.24 & 0.19 & 0.27 & 0.59 & -0.39 & -0.61 & 87.03 & 1.12 & 0.06 & 0.52 & ANV & Y \\ 
  Mean-Var (LS) & NBTW Subgraph & \eqref{noloopsabsunweighted} & central & 0.7 & 0.9 & 5.24 & 0.19 & 0.27 & 0.60 & -0.38 & -0.61 & 96.06 & 1.12 & 0.06 & 0.51 & ANV & Y \\ 
  Mean-Var (LS) & NBTW Subgraph & \eqref{loopsabsunweighted} & central & 0.7 & 0.2 & 5.22 & 0.19 & 0.28 & 0.58 & -0.41 & -0.63 & 98.13 & 1.12 & 0.05 & 0.52 & ANV & Y \\ 
  Mean-Var (LS) & NBTW Subgraph & \eqref{loopsabsunweighted} & central & 0.7 & 0.4 & 5.22 & 0.19 & 0.28 & 0.58 & -0.41 & -0.63 & 98.13 & 1.12 & 0.05 & 0.52 & ANV & Y \\ 
  Mean-Var (LS) & NBTW Subgraph & \eqref{loopsabsunweighted} & central & 0.7 & 0.8 & 5.22 & 0.19 & 0.28 & 0.58 & -0.41 & -0.63 & 98.13 & 1.12 & 0.05 & 0.52 & ANV & Y \\ 
  Mean-Var (LS) & NBTW Subgraph & \eqref{loopsabsunweighted} & central & 0.7 & 0.9 & 5.22 & 0.19 & 0.28 & 0.58 & -0.41 & -0.63 & 98.13 & 1.12 & 0.05 & 0.52 & ANV & Y \\ 
  Mean-Var (LS) & Exponential & \eqref{loopsabsunweighted} & central & 0.7 & 0.4 & 5.22 & 0.19 & 0.28 & 0.58 & -0.41 & -0.62 & 87.03 & 1.12 & 0.05 & 0.52 & ANV & Y \\ 
\hline
\end{tabular}
}
\end{sidewaystable}

\begin{sidewaystable}[h!]
\caption{SR 20 - A2}  
\bigskip
\centering 
\resizebox{\linewidth}{!}{
\begin{tabular}{llllllrrrrrrrrrrrl}
  \hline
Portfolio & Centrality & AM & C/P & TV & alpha & CR & ER & SD & SR & VaR005 & CVaR005 & MaxDDar & Omega & Sortino & UP & Graph & Shrink \\ 
\hline
Min Var (LS) & Exponential & \eqref{noloopsabsweighted} & peripheral & 0.7 & 0.2 & 4.50 & 0.16 & 0.17 & 0.80 & -0.24 & -0.38 & 43.10 & 1.16 & 0.07 & 0.54 & ANV & Y \\ 
  Mean-Var (LO) & Exponential & \eqref{noloopsabsunweighted} & peripheral & 0.6 & 0.5 & 4.35 & 0.16 & 0.16 & 0.80 & -0.25 & -0.35 & 38.63 & 1.15 & 0.07 & 0.57 & ANV & Y \\ 
  Min Var (LO) & Exponential & \eqref{noloopsabsunweighted} & peripheral & 0.6 & 0.5 & 4.07 & 0.15 & 0.15 & 0.80 & -0.22 & -0.34 & 36.80 & 1.15 & 0.07 & 0.56 & ANV & Y \\ 
  EW & Exponential & \eqref{noloopsabsweighted} & peripheral & 0.5 & 0.9 & 5.00 & 0.18 & 0.19 & 0.80 & -0.28 & -0.42 & 60.74 & 1.16 & 0.07 & 0.55 & ANV & Y \\ 
  Min Var (LS) & Exponential & \eqref{noloopsabsunweighted} & peripheral & 0.6 & 0.5 & 4.07 & 0.15 & 0.15 & 0.79 & -0.23 & -0.34 & 34.45 & 1.15 & 0.07 & 0.56 & ANV & Y \\ 
  Min Var (LO) & Exponential Subgraph & \eqref{noloopsabsunweighted} & peripheral & 0.6 & 0.1 & 4.15 & 0.15 & 0.15 & 0.79 & -0.23 & -0.35 & 56.09 & 1.15 & 0.07 & 0.56 & ANV & Y \\ 
  Mean-Var (LO) & Exponential Subgraph & \eqref{loopsabsunweighted} & peripheral & 0.6 & 0.9 & 4.48 & 0.16 & 0.17 & 0.79 & -0.25 & -0.38 & 40.07 & 1.15 & 0.07 & 0.56 & ANV & Y \\ 
  Min Var (LO) & Exponential & \eqref{noloopsabsweighted} & peripheral & 0.7 & 0.2 & 4.40 & 0.16 & 0.17 & 0.78 & -0.25 & -0.39 & 43.03 & 1.15 & 0.07 & 0.54 & ANV & Y \\ 
  Min Var (LO) & Exponential Subgraph & \eqref{noloopsabsunweighted} & peripheral & 0.6 & 0.4 & 4.11 & 0.15 & 0.16 & 0.78 & -0.23 & -0.35 & 49.46 & 1.15 & 0.07 & 0.56 & ANV & Y \\ 
  EW & Exponential & \eqref{noloopsabsunweighted} & peripheral & 0.5 & 0.5 & 4.89 & 0.17 & 0.19 & 0.78 & -0.28 & -0.43 & 70.74 & 1.15 & 0.07 & 0.54 & ANV & Y \\ 
  Min Var (LS) & Exponential Subgraph & \eqref{noloopsabsunweighted} & peripheral & 0.6 & 0.4 & 4.10 & 0.15 & 0.16 & 0.77 & -0.23 & -0.35 & 46.53 & 1.15 & 0.07 & 0.56 & ANV & Y \\ 
  Min Var (LO) & NBTW Subgraph & \eqref{noloopsabsunweighted} & peripheral & 0.5 & 0.5 & 4.26 & 0.15 & 0.16 & 0.77 & -0.24 & -0.36 & 45.62 & 1.15 & 0.07 & 0.56 & ANV & Y \\ 
  EW & NBTW Subgraph & \eqref{noloopsabsunweighted} & peripheral & 0.5 & 0.5 & 4.77 & 0.17 & 0.19 & 0.77 & -0.28 & -0.43 & 50.37 & 1.15 & 0.07 & 0.55 & ANV & Y \\ 
  Min Var (LS) & Exponential & \eqref{noloopsabsweighted} & peripheral & 0.7 & 0.1 & 4.35 & 0.16 & 0.17 & 0.77 & -0.24 & -0.38 & 48.40 & 1.15 & 0.07 & 0.54 & ANV & Y \\ 
  Min Var (LO) & Exponential & \eqref{noloopsabsunweighted} & peripheral & 0.5 & 1 & 4.12 & 0.15 & 0.16 & 0.77 & -0.24 & -0.35 & 44.93 & 1.15 & 0.07 & 0.56 & ANV & Y \\ 
  Min Var (LS) & Katz Min & \eqref{noloopsabsunweighted} & peripheral & 0.5 & NaN & 4.35 & 0.16 & 0.17 & 0.76 & -0.26 & -0.38 & 31.24 & 1.14 & 0.07 & 0.56 & ANV & Y \\ 
  Min Var (LS) & Exponential & \eqref{noloopsabsunweighted} & peripheral & 0.5 & 1 & 4.10 & 0.15 & 0.16 & 0.76 & -0.24 & -0.35 & 44.94 & 1.14 & 0.07 & 0.56 & ANV & Y \\ 
  Mean-Var (LO) & Exponential Subgraph & \eqref{noloopsabsunweighted} & peripheral & 0.6 & 0.4 & 4.31 & 0.15 & 0.17 & 0.76 & -0.25 & -0.38 & 58.19 & 1.14 & 0.07 & 0.56 & ANV & Y \\ 
  Min Var (LS) & NBTW Subgraph & \eqref{noloopsabsunweighted} & peripheral & 0.5 & 0.5 & 4.21 & 0.15 & 0.16 & 0.76 & -0.24 & -0.36 & 46.38 & 1.15 & 0.07 & 0.56 & ANV & Y \\ 
  Min Var (LO) & Exponential Subgraph & \eqref{noloopsabsunweighted} & peripheral & 0.6 & 0.8 & 4.11 & 0.15 & 0.16 & 0.76 & -0.24 & -0.36 & 43.64 & 1.14 & 0.07 & 0.55 & ANV & Y \\ 

\hline
\end{tabular}
}
\end{sidewaystable}

\begin{sidewaystable}[h!]
\caption{ER 20 - A3}  
\bigskip
\centering 
\resizebox{\linewidth}{!}{
\begin{tabular}{llllllrrrrrrrrrrrl}
  \hline
Portfolio & Centrality & AM & C/P & TV & alpha & CR & ER & SD & SR & VaR005 & CVaR005 & MaxDDar & Omega & Sortino & UP & Graph & Shrink \\ 
\hline
Mean-Var (LS) & Exponential Subgraph & \eqref{loopsnotransformunweighted} & central & 0.7 & 0.1 & 5.51 & 0.20 & 0.28 & 0.61 & -0.39 & -0.63 & 98.13 & 1.12 & 0.06 & 0.52 & ANV & N \\ 
  Mean-Var (LS) & NBTW Subgraph & \eqref{noloopsnotransformunweighted} & central & 0.7 & 0.9 & 5.42 & 0.19 & 0.27 & 0.61 & -0.38 & -0.62 & 98.13 & 1.13 & 0.06 & 0.51 & ANV & N \\ 
  Mean-Var (LS) & NBTW Subgraph & \eqref{noloopsnotransformunweighted} & central & 0.7 & 0.4 & 5.38 & 0.19 & 0.27 & 0.61 & -0.38 & -0.62 & 98.13 & 1.12 & 0.06 & 0.51 & ANV & N \\ 
  Mean-Var (LS) & Exponential Subgraph & \eqref{noloopsnotransformunweighted} & central & 0.7 & 0.1 & 5.33 & 0.19 & 0.28 & 0.59 & -0.39 & -0.63 & 98.13 & 1.12 & 0.06 & 0.51 & ANV & N \\ 
  Mean-Var (LS) & Katz Subgraph & \eqref{loopsnotransformunweighted} & central & 0.7 & 0.8 & 5.30 & 0.19 & 0.28 & 0.58 & -0.41 & -0.64 & 98.13 & 1.12 & 0.05 & 0.51 & ANV & N \\ 
  Mean-Var (LS) & NBTW Subgraph & \eqref{loopsnotransformunweighted} & central & 0.7 & 0.6 & 5.30 & 0.19 & 0.28 & 0.58 & -0.41 & -0.64 & 98.13 & 1.12 & 0.05 & 0.52 & ANV & N \\ 
  Mean-Var (LS) & Katz Subgraph & \eqref{loopsnotransformunweighted} & central & 0.7 & 0.7 & 5.27 & 0.19 & 0.28 & 0.58 & -0.40 & -0.64 & 98.13 & 1.12 & 0.05 & 0.51 & ANV & N \\ 
  Mean-Var (LS) & Exponential Subgraph & \eqref{noloopsnotransformunweighted} & central & 0.7 & 0.3 & 5.26 & 0.19 & 0.28 & 0.58 & -0.40 & -0.63 & 94.28 & 1.12 & 0.05 & 0.51 & ANV & N \\ 
  Mean-Var (LS) & NBTW Subgraph & \eqref{noloopsnotransformunweighted} & central & 0.7 & 0.5 & 5.26 & 0.19 & 0.27 & 0.59 & -0.38 & -0.62 & 98.13 & 1.12 & 0.06 & 0.51 & ANV & N \\ 
  Mean-Var (LS) & Exponential Subgraph & \eqref{loopsnotransformunweighted} & central & 0.7 & 0.2 & 5.25 & 0.19 & 0.28 & 0.58 & -0.40 & -0.62 & 94.28 & 1.12 & 0.05 & 0.52 & ANV & N \\ 
  Mean-Var (LS) & Katz Subgraph & \eqref{noloopsnotransformunweighted} & central & 0.7 & 0.8 & 5.25 & 0.19 & 0.28 & 0.57 & -0.40 & -0.64 & 98.13 & 1.12 & 0.05 & 0.51 & ANV & N \\ 
  Mean-Var (LS) & NBTW Subgraph & \eqref{noloopsnotransformunweighted} & central & 0.7 & 0.3 & 5.24 & 0.19 & 0.27 & 0.59 & -0.38 & -0.62 & 98.13 & 1.12 & 0.06 & 0.51 & ANV & N \\ 
  Mean-Var (LS) & NBTW Subgraph & \eqref{loopsnotransformunweighted} & central & 0.7 & 0.7 & 5.24 & 0.19 & 0.28 & 0.57 & -0.41 & -0.64 & 98.13 & 1.11 & 0.05 & 0.51 & ANV & N \\ 
  Mean-Var (LS) & NBTW Subgraph & \eqref{noloopsnotransformunweighted} & central & 0.7 & 0.1 & 5.23 & 0.19 & 0.27 & 0.59 & -0.38 & -0.62 & 98.13 & 1.12 & 0.06 & 0.51 & ANV & N \\ 
  Mean-Var (LS) & Katz Subgraph & \eqref{noloopsnotransformunweighted} & central & 0.7 & 0.7 & 5.21 & 0.19 & 0.28 & 0.57 & -0.41 & -0.64 & 98.13 & 1.11 & 0.05 & 0.51 & ANV & N \\ 
  EW & Exponential Subgraph & \eqref{noloopsnotransformweighted} & peripheral & 0.5 & 0.7 & 5.21 & 0.19 & 0.20 & 0.80 & -0.28 & -0.45 & 86.76 & 1.16 & 0.07 & 0.54 & ANV & N \\ 
  Mean-Var (LS) & NBTW Subgraph & \eqref{noloopsnotransformunweighted} & central & 0.7 & 0.2 & 5.21 & 0.19 & 0.28 & 0.58 & -0.38 & -0.62 & 98.13 & 1.12 & 0.06 & 0.51 & ANV & N \\ 
  Mean-Var (LS) & Exponential Subgraph & \eqref{loopsnotransformunweighted} & central & 0.7 & 0.3 & 5.19 & 0.19 & 0.28 & 0.57 & -0.40 & -0.63 & 94.28 & 1.11 & 0.05 & 0.52 & ANV & N \\ 
  Min Var (LS) & NBTW Subgraph & \eqref{noloopsnotransformunweighted} & central & 0.7 & 0.4 & 5.17 & 0.18 & 0.28 & 0.56 & -0.37 & -0.64 & 66.09 & 1.12 & 0.05 & 0.48 & ANV & N \\ 
  Mean-Var (LS) & NBTW Subgraph & \eqref{loopsnotransformunweighted} & central & 0.7 & 0.9 & 5.17 & 0.18 & 0.28 & 0.56 & -0.41 & -0.64 & 98.13 & 1.11 & 0.05 & 0.51 & ANV & N \\ 
\hline
\end{tabular}
}
\end{sidewaystable}

\begin{sidewaystable}[h!]
\caption{SR 20 - A3}  
\bigskip
\centering 
\resizebox{\linewidth}{!}{
\begin{tabular}{llllllrrrrrrrrrrrl}
  \hline
Portfolio & Centrality & AM & C/P & TV & alpha & CR & ER & SD & SR & VaR005 & CVaR005 & MaxDDar & Omega & Sortino & UP & Graph & Shrink \\ 
\hline
Min Var (LS) & Katz Min & \eqref{noloopsnotransformunweighted} & peripheral & 0.5 & NaN & 4.75 & 0.17 & 0.17 & 0.83 & -0.26 & -0.39 & 30.45 & 1.16 & 0.08 & 0.56 & ANV & N \\ 
  EW & NBTW Subgraph & \eqref{noloopsnotransformunweighted} & peripheral & 0.5 & 0.3 & 5.01 & 0.18 & 0.19 & 0.80 & -0.29 & -0.43 & 48.48 & 1.16 & 0.07 & 0.56 & ANV & N \\ 
  EW & NBTW Subgraph & \eqref{noloopsnotransformunweighted} & peripheral & 0.4 & 0.3 & 5.02 & 0.18 & 0.19 & 0.80 & -0.28 & -0.44 & 72.52 & 1.16 & 0.07 & 0.54 & ANV & N \\ 
  EW & Exponential Subgraph & \eqref{noloopsnotransformweighted} & peripheral & 0.5 & 0.7 & 5.21 & 0.19 & 0.20 & 0.80 & -0.28 & -0.45 & 86.76 & 1.16 & 0.07 & 0.54 & ANV & N \\ 
  Mean-Var (LS) & Exponential Subgraph & \eqref{loopsnotransformunweighted} & peripheral & 0.6 & 0.1 & 4.69 & 0.17 & 0.18 & 0.79 & -0.27 & -0.41 & 47.54 & 1.15 & 0.07 & 0.56 & ANV & N \\ 
  Min Var (LO) & Exponential Subgraph & \eqref{loopsnotransformunweighted} & peripheral & 0.6 & 0.1 & 4.22 & 0.15 & 0.16 & 0.78 & -0.25 & -0.36 & 36.35 & 1.15 & 0.07 & 0.56 & ANV & N \\ 
  EW & Exponential Subgraph & \eqref{loopsnotransformweighted} & peripheral & 0.5 & 0.6 & 4.96 & 0.18 & 0.19 & 0.78 & -0.28 & -0.44 & 74.48 & 1.16 & 0.07 & 0.54 & ANV & N \\ 
  Mean-Var (LO) & Exponential Subgraph & \eqref{loopsnotransformunweighted} & peripheral & 0.6 & 0.1 & 4.50 & 0.16 & 0.17 & 0.78 & -0.27 & -0.39 & 42.86 & 1.15 & 0.07 & 0.56 & ANV & N \\ 
  EW & Exponential Subgraph & \eqref{noloopsnotransformweighted} & peripheral & 0.6 & 0.5 & 4.72 & 0.17 & 0.18 & 0.77 & -0.28 & -0.41 & 42.29 & 1.15 & 0.07 & 0.56 & ANV & N \\ 
  EW & NBTW Subgraph & \eqref{noloopsnotransformunweighted} & peripheral & 0.4 & 0.7 & 4.87 & 0.17 & 0.19 & 0.77 & -0.28 & -0.44 & 71.72 & 1.15 & 0.07 & 0.54 & ANV & N \\ 
  Mean-Var (LS) & Exponential & \eqref{loopsnotransformunweighted} & peripheral & 0.6 & 0.1 & 4.48 & 0.16 & 0.17 & 0.77 & -0.26 & -0.39 & 62.95 & 1.15 & 0.07 & 0.56 & ANV & N \\ 
  EW & Exponential Subgraph & \eqref{loopsnotransformunweighted} & peripheral & 0.6 & 0.8 & 4.98 & 0.18 & 0.20 & 0.77 & -0.29 & -0.45 & 64.50 & 1.15 & 0.07 & 0.54 & ANV & N \\ 
  EW & Exponential Subgraph & \eqref{noloopsnotransformunweighted} & peripheral & 0.4 & 1 & 4.93 & 0.18 & 0.20 & 0.76 & -0.28 & -0.44 & 65.42 & 1.15 & 0.07 & 0.53 & ANV & N \\ 
  Min Var (LS) & Exponential Subgraph & \eqref{loopsnotransformweighted} & peripheral & 0.8 & 0.6 & 4.12 & 0.15 & 0.16 & 0.75 & -0.24 & -0.36 & 52.75 & 1.14 & 0.07 & 0.56 & ANV & N \\ 
  Min Var (LO) & Exponential Subgraph & \eqref{loopsnotransformweighted} & peripheral & 0.8 & 0.6 & 4.11 & 0.15 & 0.16 & 0.75 & -0.25 & -0.36 & 58.75 & 1.14 & 0.07 & 0.56 & ANV & N \\ 
  Min Var (LS) & Exponential Subgraph & \eqref{loopsnotransformunweighted} & peripheral & 0.6 & 0.1 & 4.06 & 0.14 & 0.16 & 0.75 & -0.25 & -0.36 & 38.06 & 1.14 & 0.07 & 0.56 & ANV & N \\ 
  Min Var (LO) & Katz Min & \eqref{noloopsnotransformunweighted} & peripheral & 0.5 & NaN & 4.40 & 0.16 & 0.18 & 0.75 & -0.27 & -0.40 & 39.83 & 1.14 & 0.07 & 0.55 & ANV & N \\ 
  EW & Exponential Subgraph & \eqref{noloopsnotransformweighted} & peripheral & 0.5 & 0.3 & 4.44 & 0.16 & 0.18 & 0.75 & -0.26 & -0.40 & 41.84 & 1.14 & 0.07 & 0.55 & ANV & N \\ 
  EW & NBTW Subgraph & \eqref{loopsnotransformweighted} & peripheral & 0.5 & 0.2 & 4.64 & 0.17 & 0.19 & 0.74 & -0.28 & -0.43 & 64.35 & 1.14 & 0.07 & 0.55 & ANV & N \\ 
  EW & NBTW Subgraph & \eqref{noloopsnotransformunweighted} & peripheral & 0.5 & 0.2 & 4.46 & 0.16 & 0.18 & 0.74 & -0.28 & -0.41 & 68.41 & 1.14 & 0.07 & 0.55 & ANV & N \\ 
\hline
\end{tabular}
}
\end{sidewaystable}

\begin{sidewaystable}[h!]
\caption{ER 20 - A4}  
\bigskip
\centering 
\resizebox{\linewidth}{!}{
\begin{tabular}{llllllrrrrrrrrrrrl}
  \hline
Portfolio & Centrality & AM & C/P & TV & alpha & CR & ER & SD & SR & VaR005 & CVaR005 & MaxDDar & Omega & Sortino & UP & Graph & Shrink \\ 
\hline
Mean-Var (LS) & Exponential Subgraph & \eqref{loopsnotransformunweighted} & central & 0.7 & 0.1 & 5.57 & 0.20 & 0.28 & 0.63 & -0.40 & -0.62 & 98.13 & 1.13 & 0.06 & 0.52 & ANV & Y \\ 
  Mean-Var (LS) & Katz Subgraph & \eqref{loopsnotransformunweighted} & central & 0.7 & 0.7 & 5.48 & 0.20 & 0.28 & 0.61 & -0.40 & -0.63 & 98.13 & 1.12 & 0.06 & 0.52 & ANV & Y \\ 
  Mean-Var (LS) & Exponential Subgraph & \eqref{noloopsnotransformunweighted} & central & 0.7 & 0.1 & 5.47 & 0.20 & 0.28 & 0.61 & -0.40 & -0.62 & 98.13 & 1.12 & 0.06 & 0.52 & ANV & Y \\ 
  Mean-Var (LS) & Exponential Subgraph & \eqref{loopsnotransformunweighted} & central & 0.7 & 0.2 & 5.43 & 0.19 & 0.28 & 0.61 & -0.40 & -0.61 & 88.86 & 1.12 & 0.06 & 0.52 & ANV & Y \\ 
  Mean-Var (LS) & Exponential & \eqref{noloopsnotransformunweighted} & central & 0.7 & 0.2 & 5.38 & 0.19 & 0.28 & 0.60 & -0.40 & -0.62 & 88.86 & 1.12 & 0.06 & 0.52 & ANV & Y \\ 
  Mean-Var (LS) & Exponential & \eqref{loopsnotransformunweighted} & central & 0.7 & 0.2 & 5.35 & 0.19 & 0.28 & 0.60 & -0.40 & -0.62 & 88.86 & 1.12 & 0.06 & 0.52 & ANV & Y \\ 
  Mean-Var (LS) & Exponential Subgraph & \eqref{noloopsnotransformunweighted} & central & 0.7 & 0.2 & 5.33 & 0.19 & 0.27 & 0.60 & -0.40 & -0.61 & 88.86 & 1.12 & 0.06 & 0.52 & ANV & Y \\ 
  Mean-Var (LS) & Exponential Subgraph & \eqref{noloopsnotransformunweighted} & central & 0.7 & 0.3 & 5.33 & 0.19 & 0.27 & 0.60 & -0.39 & -0.61 & 88.86 & 1.12 & 0.06 & 0.52 & ANV & Y \\ 
  Mean-Var (LS) & Katz Subgraph & \eqref{noloopsnotransformunweighted} & central & 0.7 & 0.7 & 5.32 & 0.19 & 0.28 & 0.59 & -0.41 & -0.63 & 98.13 & 1.12 & 0.06 & 0.52 & ANV & Y \\ 
  Mean-Var (LS) & NBTW Subgraph & \eqref{loopsnotransformunweighted} & central & 0.7 & 0.6 & 5.32 & 0.19 & 0.28 & 0.59 & -0.41 & -0.63 & 98.13 & 1.12 & 0.06 & 0.52 & ANV & Y \\ 
  Mean-Var (LS) & NBTW Subgraph & \eqref{loopsnotransformunweighted} & central & 0.7 & 0.7 & 5.30 & 0.19 & 0.28 & 0.59 & -0.41 & -0.63 & 98.13 & 1.12 & 0.06 & 0.52 & ANV & Y \\ 
  Mean-Var (LS) & Exponential Subgraph & \eqref{loopsnotransformunweighted} & central & 0.7 & 0.3 & 5.29 & 0.19 & 0.28 & 0.59 & -0.40 & -0.61 & 88.86 & 1.12 & 0.06 & 0.52 & ANV & Y \\ 
  Mean-Var (LS) & Katz Subgraph & \eqref{noloopsnotransformunweighted} & central & 0.7 & 0.8 & 5.28 & 0.19 & 0.28 & 0.58 & -0.40 & -0.62 & 96.06 & 1.12 & 0.06 & 0.52 & ANV & Y \\ 
  Mean-Var (LS) & Katz Subgraph & \eqref{loopsnotransformunweighted} & central & 0.7 & 0.6 & 5.26 & 0.19 & 0.28 & 0.58 & -0.40 & -0.62 & 98.13 & 1.12 & 0.06 & 0.52 & ANV & Y \\ 
  Mean-Var (LS) & Exponential Subgraph & \eqref{loopsnotransformunweighted} & central & 0.7 & 0.4 & 5.25 & 0.19 & 0.28 & 0.59 & -0.40 & -0.61 & 88.86 & 1.12 & 0.06 & 0.52 & ANV & Y \\ 
  Mean-Var (LS) & NBTW Subgraph & \eqref{noloopsnotransformunweighted} & central & 0.7 & 0.9 & 5.24 & 0.19 & 0.27 & 0.60 & -0.38 & -0.61 & 96.06 & 1.12 & 0.06 & 0.51 & ANV & Y \\ 
  Mean-Var (LS) & Katz & \eqref{noloopsnotransformunweighted} & central & 0.7 & 0.8 & 5.23 & 0.19 & 0.28 & 0.58 & -0.41 & -0.63 & 98.13 & 1.12 & 0.05 & 0.52 & ANV & Y \\ 
  Mean-Var (LS) & Exponential & \eqref{noloopsnotransformunweighted} & central & 0.7 & 0.3 & 5.23 & 0.19 & 0.28 & 0.58 & -0.41 & -0.62 & 88.86 & 1.12 & 0.05 & 0.52 & ANV & Y \\ 
  Mean-Var (LS) & NBTW Subgraph & \eqref{loopsnotransformunweighted} & central & 0.7 & 0.2 & 5.22 & 0.19 & 0.28 & 0.58 & -0.41 & -0.63 & 98.13 & 1.12 & 0.05 & 0.52 & ANV & Y \\ 
  Mean-Var (LS) & NBTW Subgraph & \eqref{loopsnotransformunweighted} & central & 0.7 & 0.4 & 5.22 & 0.19 & 0.28 & 0.58 & -0.41 & -0.63 & 98.13 & 1.12 & 0.05 & 0.52 & ANV & Y \\ 
\hline
\end{tabular}
}
\end{sidewaystable}

\begin{sidewaystable}[h!]
\caption{SR 20 - A4}  
\bigskip
\centering 
\resizebox{\linewidth}{!}{
\begin{tabular}{llllllrrrrrrrrrrrl}
  \hline
Portfolio & Centrality & AM & C/P & TV & alpha & CR & ER & SD & SR & VaR005 & CVaR005 & MaxDDar & Omega & Sortino & UP & Graph & Shrink \\ 
\hline
EW & Exponential & \eqref{noloopsnotransformweighted} & peripheral & 0.5 & 0.9 & 4.91 & 0.18 & 0.18 & 0.83 & -0.27 & -0.40 & 79.49 & 1.16 & 0.08 & 0.55 & ANV & Y \\ 
  EW & Exponential Subgraph & \eqref{noloopsnotransformweighted} & peripheral & 0.5 & 1 & 5.05 & 0.18 & 0.19 & 0.82 & -0.28 & -0.42 & 64.31 & 1.16 & 0.08 & 0.55 & ANV & Y \\ 
  EW & Exponential & \eqref{loopsnotransformweighted} & peripheral & 0.5 & 1 & 5.20 & 0.19 & 0.20 & 0.82 & -0.29 & -0.43 & 70.59 & 1.16 & 0.08 & 0.55 & ANV & Y \\ 
  EW & Exponential Subgraph & \eqref{noloopsnotransformunweighted} & peripheral & 0.5 & 0.5 & 4.94 & 0.18 & 0.19 & 0.81 & -0.28 & -0.42 & 75.46 & 1.15 & 0.07 & 0.55 & ANV & Y \\ 
  EW & Exponential & \eqref{noloopsnotransformunweighted} & peripheral & 0.5 & 0.5 & 4.99 & 0.18 & 0.19 & 0.81 & -0.28 & -0.41 & 59.33 & 1.16 & 0.07 & 0.56 & ANV & Y \\ 
  EW & Exponential Subgraph & \eqref{noloopsnotransformweighted} & peripheral & 0.4 & 0.8 & 5.01 & 0.18 & 0.19 & 0.79 & -0.28 & -0.42 & 74.12 & 1.16 & 0.07 & 0.54 & ANV & Y \\ 
  EW & Exponential Subgraph & \eqref{noloopsnotransformweighted} & peripheral & 0.5 & 0.5 & 4.91 & 0.18 & 0.19 & 0.79 & -0.28 & -0.42 & 68.83 & 1.15 & 0.07 & 0.55 & ANV & Y \\ 
  EW & Exponential Subgraph & \eqref{noloopsnotransformweighted} & peripheral & 0.5 & 0.7 & 4.75 & 0.17 & 0.18 & 0.79 & -0.28 & -0.41 & 67.95 & 1.15 & 0.07 & 0.56 & ANV & Y \\ 
  EW & Exponential Subgraph & \eqref{loopsnotransformunweighted} & peripheral & 0.5 & 1 & 4.62 & 0.16 & 0.18 & 0.79 & -0.26 & -0.40 & 50.94 & 1.15 & 0.07 & 0.55 & ANV & Y \\ 
  EW & Exponential Subgraph & \eqref{loopsnotransformweighted} & peripheral & 0.5 & 0.2 & 4.92 & 0.18 & 0.19 & 0.78 & -0.28 & -0.43 & 67.36 & 1.15 & 0.07 & 0.55 & ANV & Y \\ 
  EW & Exponential Subgraph & \eqref{loopsnotransformunweighted} & peripheral & 0.5 & 0.3 & 5.03 & 0.18 & 0.20 & 0.77 & -0.28 & -0.44 & 76.37 & 1.16 & 0.07 & 0.54 & ANV & Y \\ 
  Min Var (LO) & NBTW Subgraph & \eqref{noloopsnotransformunweighted} & peripheral & 0.5 & 0.5 & 4.26 & 0.15 & 0.16 & 0.77 & -0.24 & -0.36 & 45.62 & 1.15 & 0.07 & 0.56 & ANV & Y \\ 
  EW & NBTW Subgraph & \eqref{noloopsnotransformunweighted} & peripheral & 0.5 & 0.5 & 4.77 & 0.17 & 0.19 & 0.77 & -0.28 & -0.43 & 50.37 & 1.15 & 0.07 & 0.55 & ANV & Y \\ 
  Min Var (LS) & Katz Min & \eqref{noloopsnotransformunweighted} & peripheral & 0.5 & NaN & 4.53 & 0.16 & 0.18 & 0.77 & -0.27 & -0.40 & 30.45 & 1.15 & 0.07 & 0.56 & ANV & Y \\ 
  Min Var (LO) & Exponential & \eqref{noloopsnotransformweighted} & peripheral & 0.5 & 0.9 & 4.09 & 0.15 & 0.16 & 0.76 & -0.23 & -0.35 & 52.52 & 1.15 & 0.07 & 0.55 & ANV & Y \\ 
  EW & Exponential Subgraph & \eqref{noloopsnotransformweighted} & peripheral & 0.4 & 0.1 & 4.82 & 0.17 & 0.19 & 0.76 & -0.27 & -0.42 & 75.48 & 1.15 & 0.07 & 0.54 & ANV & Y \\ 
  Min Var (LS) & NBTW Subgraph & \eqref{noloopsnotransformunweighted} & peripheral & 0.5 & 0.5 & 4.21 & 0.15 & 0.16 & 0.76 & -0.24 & -0.36 & 46.38 & 1.15 & 0.07 & 0.56 & ANV & Y \\ 
  EW & Exponential Subgraph & \eqref{noloopsnotransformunweighted} & peripheral & 0.5 & 0.8 & 4.65 & 0.17 & 0.19 & 0.76 & -0.28 & -0.42 & 74.91 & 1.15 & 0.07 & 0.55 & ANV & Y \\ 
  Mean-Var (LO) & Exponential Subgraph & \eqref{noloopsnotransformunweighted} & peripheral & 0.6 & 0.9 & 4.41 & 0.16 & 0.17 & 0.76 & -0.26 & -0.38 & 36.49 & 1.14 & 0.07 & 0.56 & ANV & Y \\ 
  EW & Exponential & \eqref{noloopsnotransformunweighted} & peripheral & 0.5 & 1 & 4.57 & 0.16 & 0.18 & 0.76 & -0.27 & -0.41 & 64.84 & 1.15 & 0.07 & 0.54 & ANV & Y \\ 
\hline
\end{tabular}
}
\end{sidewaystable}

\section{Conclusions}
Our findings suggest that the following factors contribute to achieving high Sharpe Ratios (SR) in the portfolios:
\begin{itemize}
    \item Centrality measures such as Exponential, Katz Min, NBTW Subgraph and Exponential Subgraph.
    \item Constructing portfolios predominantly with peripheral stocks. 
    \item Employing equally weighted portfolios and minimum variance strategies. 
    \item Among the adjacency matrix options, the configurations (7) (no loops, no transform and weighted graphs), (3) (no loops, no transform and unweighted graphs) and (4) (no loops, absolute value transform and unweighted graphs) tend to contribute to portfolios yielding the highest SR.
    \item The threshold values most commonly used are $0.5$ and $0.6$.
    \item Shrinkage seems to depend on the original transformation applied to the correlation matrix and does not appear always as a common practice among portfolios with the highest SR.
\end{itemize}
The following factors contribute to portfolios with high expected return:
\begin{itemize}
\item Employing subgraph centrality measures such as Exponential Subgraph, NBTW Subgraph, and Katz Subgraph.
\item Constructing portfolios predominantly with central stocks.
\item Utilising mean-variance (long-short) portfolios.
\item A threshold value of 0.7 is used almost always in achieving high ER.
\item Shrinkage seems again to depend on the original transformation applied to the correlation matrix and does not appear always as a common practice among portfolios with the highest ER.
\item Among the adjacency matrix options, configurations (2) (loops, absolute value transform, and unweighted graphs), (4) (no loops, absolute value transform, and unweighted graphs), (1) (loops, no transform, unweighted) and (3) (no loops, no transform and unweighted graphs) exhibit the highest frequency.
\end{itemize}

Summarising our findings
\begin{itemize}
\item We support the claim already appearing in the literature that periphery outperforms centre when the metric is SR.
\item We have tested a much wider range of centrality measures, including the fairly new NBTW ones which perform very well.
\item We have proposed the use of subgraph centralities (Exponential, Katz and NBTW) which appear to do quite well as a whole and individually.
\item We have proposed the use of threshold instead of other more sophisticated filtering techniques. We have found that in spite of its simplicity threshold actually does better and have depicted the most used threshold values. 
\item How to choose $\alpha$ appears to be still rather elusive, but we have shown the most frequent values of $\alpha$ for each centrality and transformation used.
\item Equally weighted portfolios, in spite of their simplicity, outperform minimum variance or mean variance in most of the cases.
\item The methods proposed outperform other previously proposed methods such as the MST.

\end{itemize}

\end{document}